\documentclass[aps,prd,twocolumn,superscriptaddress]{revtex4}
\usepackage{epsfig,epsf}
\usepackage{amsmath}
\usepackage{amsthm}
\usepackage{amsfonts}
\usepackage{amssymb}
\usepackage{dsfont}
\usepackage{multirow}
\usepackage{appendix}
\usepackage{slashed}
\usepackage[active]{srcltx}
\usepackage{psfrag}

\begin{document}

\title{Spectroscopic parameters and decays of the resonance $Z_b(10610)$}
\date{\today}
\author{S.~S.~Agaev}
\affiliation{Institute for Physical Problems, Baku State University, Az--1148 Baku,
Azerbaijan}
\author{K.~Azizi}
\affiliation{Department of Physics, Do\v{g}u\c{s} University, Acibadem-Kadik\"{o}y, 34722
Istanbul, Turkey}
\affiliation{School of Physics, Institute for Research in Fundamental Sciences (IPM),
P.~O.~Box 19395-5531, Tehran, Iran}
\author{H.~Sundu}
\affiliation{Department of Physics, Kocaeli University, 41380 Izmit, Turkey}

\begin{abstract}
The resonance $Z_b(10610)$ is investigated as the diquark-antidiquark $%
Z_b=[bu][\overline b \overline d]$ state with spin-parity $J^{P}=1^{+}$. The
mass and current coupling of the resonance $Z_b(10610)$ are evaluated using
QCD two-point sum rule and taking into account the vacuum condensates up to
ten dimensions. We study the vertices $Z_b\Upsilon(nS)\pi\ (n=1,2,3)$ by
applying the QCD light-cone sum rule to compute the corresponding strong
couplings $g_{Z_b\Upsilon(nS)\pi}$ and widths of the decays $Z_b \to
\Upsilon(nS)\pi$. We explore also the vertices $Z_b h_{b}(mP)\pi\ (m=1,2)$
and calculate the couplings $g_{Z_b h_{b}(mP)\pi}$ and width of the decay
channels $Z_b \to h_{b}(mP)\pi$. To this end, we calculate the mass and
decay constants of the $h_b(1P)$ and $h_b(2P) $ mesons. The results obtained
are compared with experimental data of the Belle Collaboration.
\end{abstract}

\maketitle

\section{Introduction}

Discovery of the charged resonances which cannot be explained as $\bar{c}c$
or $\bar{b}b$ states has opened a new page in physics of exotic multi-quark
systems. The first tetraquarks of this family are $Z^{\pm }(4430)$ states
which were observed by the Belle Collaboration in $B$ meson decays $%
B\rightarrow K\psi ^{\prime }\pi ^{\pm }$ as resonances in the $\psi
^{\prime }\pi ^{\pm }$ invariant mass distributions \cite{Choi:2007wga}. The
mass and width of these states were repeatedly measured and refined.
Recently, the LHCb Collaboration confirmed existence of the $Z^{-}(4430)$
structure in the decay $B^{0}\rightarrow K^{+}\psi ^{\prime }\pi ^{-}$ and
unambiguously determined that its spin-parity is $J^{P}=1^{+}$ \cite%
{Aaij:2014jqa,Aaij:2015zxa}. They also measured the mass and width of $%
Z^{-}(4430)$ resonance and updated the existing experimental data. Two
charmonium-like resonances $Z_{1}(4050)$ and $Z_{2}(4250)$ were discovered
by the Belle Collaboration in the decay $\bar{B}^{0}\rightarrow K^{-}\pi
^{+}\chi _{c1}$ which emerged as broad peaks in the $\chi _{c1}\pi $
invariant mass distribution \cite{Mizuk:2008me}.

Famous members of the charged tetraquark family $Z_{c}^{\pm }(3900)$ were
observed by the BESIII Collaboration in the process $e^{+}e^{-}\rightarrow
J/\psi \pi ^{+}\pi ^{-}$ as resonances with $J^{P}=1^{+}$ in the $J/\psi \pi
^{\pm }$ mass distribution \cite{Ablikim:2013mio}. The charged state $%
Z_{c}(4020)$ was also found by the BESIII Collaboration in two different
processes $e^{+}e^{-}\rightarrow h_{c}\pi ^{+}\pi ^{-}$ and $%
e^{+}e^{-}\rightarrow (D^{\star }\bar{D}^{\star })^{\pm }\pi ^{\mp }$ (see,
Refs.\ \cite{Ablikim:2013wzq,Ablikim:2013emm}).

There is another charged state, namely $Z_{c}(4200)$ resonance which was
detected and announced by Belle \cite{Chilikin:2014bkk}. All aforementioned
resonances belong to the class of the charmonium-like tetraquarks, and
contain a $\bar{c}c$ pair and light quarks (antiquarks). They were mainly
interpreted as diquark-antidiquark systems or bound states of $D$ and/or $%
D^{\star }$ mesons.

It is remarkable, that $b$-counterparts of the charmonium-like states, i.e.
charged resonances composed of a $\bar{b}b$ pair and light quarks were
found, as well. Thus, the Belle Collaboration discovered the resonances $%
Z_{b}(10610)$ and $Z_{b}(10650)$ (hereafter, $Z_{b}$ and $Z_{b}^{\prime }$,
respectively) in the decays $\Upsilon (5S)\rightarrow \Upsilon (nS)\pi
^{+}\pi ^{-},\ n=1,2,3$ and $\Upsilon (5S)\rightarrow h_{b}(mP)\pi ^{+}\pi
^{-},\ m=1,2$ \cite{Belle:2011aa,Garmash:2014dhx}. These two states with
favored spin-parity $J^{P}=1^{+}$ appear as resonances in the $\Upsilon
(nS)\pi ^{\pm }$ and $h_{b}(mP)\pi ^{\pm }$ mass distributions. The masses
of the $Z_{b}$ and $Z_{b}^{\prime }$ resonances are
\begin{eqnarray}
m &=&(10607.2\pm 2.0)\ \mathrm{MeV},  \notag \\
m^{\prime } &=&(10652.2\pm 1.5)\ \mathrm{MeV},  \label{eq:ExpMass}
\end{eqnarray}%
respectively. The width of the $Z_{b}$ state averaged over five decay
channels equals to $\Gamma =(18.4\pm 2.4)\ \mathrm{MeV}$, whereas the
average width of $Z_{b}^{\prime }$ is $\Gamma ^{\prime }=(11.5\pm 2.2)\
\mathrm{MeV}$. Recently, the dominant decay channel of $Z_{b}$ , namely $%
Z_{b}\rightarrow B^{+}\bar{B}^{\star 0}+\bar{B}^{0}B^{\star +}$ process was
also observed \cite{Garmash:2015rfd}. In this work fractions of different
channels of $Z_{b}$ and $Z_{b}^{\prime }$ resonances were reported, as well.
Further information on experimental status of the $Z_{b}$ and $Z_{b}^{\prime
}$ states and other heavy exotic mesons and baryons can be found in Ref.\
\cite{Olsen:2017bmm}.

An existence of hidden-bottom states, i.e. of the $Z_b$ resonances were
foreseen before their experimental observation. Thus, in Ref.\ \cite%
{Karliner:2008rc} authors suggested to look for the diquark-antidiquarks
with $b\bar b u \bar d$ content as peaks in the invariant mass of $%
\Upsilon(1S)\pi$ and $\Upsilon(2S)\pi$ systems. The existence of the
molecular state $B^{\star}\bar B$ was predicted in Ref.\ \cite{Liu:2008tn}.

After discovery of the $Z_{b}$ resonances theoretical studies of the charged
hidden-bottom states became more intensive and fruitful. In fact, works
devoted to the structures and decay channels of the $Z_{b}$ states encompass
all existing models and computational schemes suitable to study the
multi-quark systems. Thus, in Refs.\ \cite{Bondar:2011ev,Voloshin:2011qa}
the spectroscopic and decay properties of $Z_{b}$ and $Z_{b}^{\prime }$ were
explored using the heavy quark symmetry by modeling them as $J=1$ $S$-wave
molecular states $B^{\star }\bar{B}-B\bar{B}^{\star }$ and $B^{\star }\bar{B}%
^{\star }$, respectively. The existence of similar states with quantum
numbers $0^{+},\ 1^{+},\ 2^{+}$ were predicted, as well. The
diquark-antidiquark interpretation of the $Z_{b}$ states were proposed in
Refs.\ \cite{Ali:2011ug,Ali:2014dva}. It was demonstrated that Belle results
on the decays $\Upsilon (5S)\rightarrow \Upsilon (nS)\pi ^{+}\pi ^{-}$ and $%
\Upsilon (5S)\rightarrow h_{b}(mP)\pi ^{+}\pi ^{-}$ support $Z_{b}$
resonances as diquark-antidiquark states. This analysis is based on a scheme
for the spin-spin quark interactions inside diquarks originally suggested
and successfully used to explore hidden-charm resonances \cite{Maiani:2014}.

The $Z_{b}$ resonance was considered in Ref.\ \cite{Zhang:2011jja} as a $%
B^{\star }\bar{B}$ molecular state, where its mass was computed in the
context of QCD sum rule method. The prediction for the mass $m_{B^{\star }%
\bar{B}}=10.54\pm 0.22\ \mathrm{GeV}$ obtained there, allowed authors to
conclude that $Z_{b}$ could be a $B^{\star }\bar{B}$ molecular state. The
similar conclusions were also made in the framework of the chiral quark
model. Indeed, in Ref.\ \cite{Yang:2011rp} the $B\bar{B}^{\star }$ and $%
B^{\star }\bar{B}^{\star }$ bound states with $J^{PC}=1^{+-}$ were studied
in the chiral quark model, and found as good candidates for $Z_{b}$ and $%
Z_{b}^{\prime }$ resonances. Moreover, existence of molecular states $B\bar{B%
}^{\star }$ with $J^{PC}=1^{++}$, and $B^{\star }\bar{B}^{\star }$ with $%
J^{PC}=0^{++},\ 2^{++}$ were predicted. Explorations performed using the one
boson-exchange model also led to the molecular interpretations of the $Z_{b}$
and $Z_{b}^{\prime }$ resonances \cite{Sun:2011uh}. However, analysis
carried out in the framework of the Bete-Salpeter approach demonstrated that
two heavy mesons can form an isospin singlet bound state but cannot form an
isotriplet compound. Hence, the $Z_{b}$ resonance presumably is a
diquark-antidiquark, but not a molecular state \cite{Ke:2012gm}.

The both diquark-antidiquark and molecular pictures for internal
organization of $Z_{b}$ and $Z_{b}^{\prime }$ within QCD sum rules method
were examined in Ref.\ \cite{Cui:2011fj}. In this work the authors
constructed different interpolating currents with $I^{G}J^{P}=1^{+}1^{+}$ to
explore the $Z_{b}$ and $Z_{b}^{\prime }$ states and evaluate their masses.
Among alternative interpretations of the $Z_{b}$ states it is worth noting
Refs.\ \cite{Bugg:2011jr} and \cite{Danilkin:2011sh}, where the peaks
observed by the Belle Collaboration were explained as cusp and coupling
channel effects, respectively.

Theoretical works that address problems of the $Z_{b}$ states are numerous
(see, Refs. \cite{Chen:2011zv,Chen:2011pv,Cleven:2011gp,Cleven:2013sq,
Mehen:2013mva,Wang:2013daa,Wang:2014gwa,Dong:2012hc,Chen:2015ata,Kang:2016ezb}%
). Analysis of these and other investigations can be found in the recent
review papers \cite{Esposito:2016noz,Ali:2017jda}.

As is seen, theoretical status of the resonances $Z_{b}$ and $Z_{b}^{\prime
} $ remains controversial and deserves further and detailed explorations. In
the present work we are going to calculate the spectroscopic parameters of $%
Z_{b}=[bu][\overline b \overline d]$ state by assuming that it is a
tetraquark state with diquark-antidiquark structure and positive charge. We
use QCD two-point sum rules to evaluate its mass and current coupling by
taking into account vacuum condensates up to ten dimensions. We also
investigate five observed decay channels of $Z_{b}$ resonance employing QCD
sum rules on the light-cone. As a byproduct, we derive the mass and decay
constant of $h_{b}(mP),\ m=1,2$ mesons.

This work has the following structure: In Sec.\ \ref{sec:MassCoupl} we
calculate the mass and current coupling of the $Z_{b}$ resonance. In Sec.\ %
\ref{sec:Decays1} we analyze the decay channels $Z_{b}\rightarrow \Upsilon
(nS)\pi ,\ n=1,\ 2,\ 3$ and calculate their widths. Section \ref{sec:Decays2}
is devoted to investigation of the decay modes $Z_{b}\rightarrow
h_{b}(mP)\pi ,\ m=1,\ 2$ and consists of two subsections. In the first
subsection we calculate the mass and decay constant of the $h_{b}(1P)$ and $%
h_{b}(2P)$ mesons. To this end, we employ the two-point sum rule approach by
including into analysis condensates up to eight dimensions. In the next
subsection using parameters of the $h_{b}(mP)$ mesons we evaluate width of
decays under investigation. The last section is reserved for analysis of the
obtained results and discussion of possible interpretations of $Z_{b}$
resonance.


\section{Mass and current coupling of the $Z_{b}$ state: QCD two-point sum
rule predictions}

\label{sec:MassCoupl}
In this section we derive QCD sum rules to calculate the mass and current
coupling of the $Z_{b}$ state by suggesting that it has a
diquark-antidiquark structure with quantum numbers $I^{G}J^{P}=1^{+}1^{+}$.
\ To this end, we begin from the two-point correlation function
\begin{equation}
\Pi _{\mu \nu }(p)=i\int d^{4}xe^{ipx}\langle 0|\mathcal{T}\{J_{\mu
}^{Z_{b}}(x)J_{\nu }^{Z_{b}\dag }(0)\}|0\rangle ,  \label{eq:CorrF1}
\end{equation}%
where $J_{\mu }^{Z_{b}}(x)$ is the interpolating current for the $Z_{b}$
state with required quark content and quantum numbers.

It is possible to construct various currents to interpolate the $Z_{b}$ and $%
Z_{b}^{\prime }$ resonances. One of them is $[ub]_{S=0}[\overline{d}%
\overline{b}]_{S=1}-$ $[ub]_{S=1}[\overline{d}\overline{b}]_{S=0}$ type
diquark-antidiquark current that is used to consider $Z_{b}$ state
\begin{eqnarray}
&&J_{\mu }^{Z_{b}}(x)=\frac{i\epsilon \tilde{\epsilon}}{\sqrt{2}}\left\{ %
\left[ u_{a}^{T}(x)C\gamma _{5}b_{b}(x)\right] \left[ \overline{d}%
_{d}(x)\gamma _{\mu }C\overline{b}_{e}^{T}(x)\right] \right.   \notag \\
&&\left. -\left[ u_{a}^{T}(x)C\gamma _{\mu }b_{b}(x)\right] \left[ \overline{%
d}_{d}(x)\gamma _{5}C\overline{b}_{e}^{T}(x)\right] \right\} .
\label{eq:CDiq}
\end{eqnarray}%
The current for $Z_{b}^{\prime }$ can be defined in the form%
\begin{eqnarray}
&&J_{\mu }^{Z_{b}^{\prime }}(x)=\frac{\epsilon \tilde{\epsilon}}{\sqrt{2}}%
\varepsilon _{\mu \nu \alpha \beta }\left[ u_{a}^{T}(x)C\gamma ^{\nu
}b_{b}(x)\right]   \notag \\
&&\times D^{\alpha }\left[ \overline{d}_{d}(x)\gamma ^{\beta }C\overline{b}%
_{e}^{T}(x)\right] ,  \label{eq:CDiqB}
\end{eqnarray}%
where $D^{\alpha }=\partial ^{\alpha }-ig_{s}A^{\alpha }(x)$ \cite%
{Cui:2011fj}. In Eqs.\ (\ref{eq:CDiq}) and \ (\ref{eq:CDiqB}) we have
introduced the notations $\epsilon =\epsilon _{abc}$ and $\tilde{\epsilon}%
=\epsilon _{dec}$. In above expressions $a,b,c,d$ and $e$ are color indices,
and $C$ is the charge conjugation matrix.

By choosing different currents to interpolate the $Z_{b}$ and $Z_{b}^{\prime
}$ resonances one treats both of them as ground-state particles in
corresponding sum rules. We also follow this approach and use the current $%
J_{\mu }^{Z_{b}}(x)$ to calculate the mass and current coupling of the $%
Z_{b} $ state. To find the QCD sum rules we first have to calculate the
correlation function in terms of the physical degrees of freedom. To this
end, we saturate $\Pi _{\mu \nu }(p)$ with a complete set of states with
quantum numbers of $Z_{b}$ resonance and perform in Eq.\ (\ref{eq:CorrF1})
integration over $x$ to get
\begin{equation*}
\Pi _{\mu \nu }^{\mathrm{Phys}}(p)=\frac{\langle 0|J_{\mu
}^{Z_{b}}|Z_{b}(p)\rangle \langle Z_{b}(p)|J_{\nu }^{Z_{b}\dagger }|0\rangle
}{m_{Z_{b}}^{2}-p^{2}}+...,
\end{equation*}%
where $m_{Z_{b}}$ is the mass of the $Z_{b}$ state, and dots indicate
contributions of higher resonances and continuum states. We define the
current coupling $f_{Z_{b}}$ through the matrix element
\begin{equation}
\langle 0|J_{\mu }^{Z_{b}}|Z_{b}(p)\rangle =f_{Z_{b}}m_{Z_{b}}\varepsilon
_{\mu },  \label{eq:Res}
\end{equation}%
with $\varepsilon _{\mu }$ being the polarization vector of $Z_{b}$ state.
Then in terms of $m_{Z_{b}}$ and $f_{Z_{b}}$, the correlation function can
be written in the following form
\begin{equation}
\Pi _{\mu \nu }^{\mathrm{Phys}}(p)=\frac{m_{Z_{b}}^{2}f_{Z_{b}}^{2}}{%
m_{Z_{b}}^{2}-p^{2}}\left( -g_{\mu \nu }+\frac{p_{\mu }p_{\nu }}{%
m_{Z_{b}}^{2}}\right) +\ldots  \label{eq:CorM}
\end{equation}%
The Borel transformation applied to Eq.\ (\ref{eq:CorM}) gives%
\begin{equation}
\mathcal{B}\Pi _{\mu \nu }^{\mathrm{Phys}%
}(p)=m_{Z_{b}}^{2}f_{Z_{b}}^{2}e^{-m_{Z_{b}}^{2}/M^{2}}\left( -g_{\mu \nu }+%
\frac{p_{\mu }p_{\nu }}{m_{Z_{b}}^{2}}\right) +\ldots
\end{equation}

At the next stage we derive the theoretical expression for the correlation
function $\Pi _{\mu \nu }^{\mathrm{QCD}}(p)$ in terms of the quark-gluon
degrees of freedom. It can be determined using the interpolating current $%
J_{\mu }^{Z_{b}}$ and quark propagators. After contracting in Eq.\ (\ref%
{eq:CorrF1}) the $b$-quark and light quark fields we get
\begin{eqnarray}
&&\Pi _{\mu \nu }^{\mathrm{QCD}}(p)=-\frac{i}{2}\int d^{4}xe^{ipx}\epsilon
\tilde{\epsilon}\epsilon ^{\prime }\tilde{\epsilon}^{\prime }\left\{ \mathrm{%
Tr}\left[ \gamma _{5}\widetilde{S}_{u}^{aa^{\prime }}(x)\right. \right.
\notag \\
&&\left. \times \gamma _{5}S_{b}^{bb^{\prime }}(x)\right] \mathrm{Tr}\left[
\gamma _{\mu }\widetilde{S}_{b}^{e^{\prime }e}(-x)\gamma _{\nu
}S_{d}^{d^{\prime }d}(-x)\right]  \notag \\
&&-\mathrm{Tr}\left[ \gamma _{\mu }\widetilde{S}_{b}^{e^{\prime
}e}(-x)\gamma _{5}S_{d}^{d^{\prime }d}(-x)\right] \mathrm{Tr}\left[ \gamma
_{\nu }\widetilde{S}_{u}^{aa^{\prime }}(x)\right.  \notag \\
&&\times \left. \gamma _{5}S_{b}^{bb^{\prime }}(x)\right] -\mathrm{Tr}\left[
\gamma _{5}\widetilde{S}_{u}^{a^{\prime }a}(x)\gamma _{\mu }S_{b}^{b^{\prime
}b}(x)\right]  \notag \\
&&\times \mathrm{Tr}\left[ \gamma _{5}\widetilde{S}_{b}^{e^{\prime
}e}(-x)\gamma _{\nu }S_{d}^{d^{\prime }d}(-x)\right] +\mathrm{Tr}\left[
\gamma _{\nu }\widetilde{S}_{u}^{aa^{\prime }}(x)\right.  \notag \\
&&\left. \times \left. \gamma _{\mu }S_{b}^{bb^{\prime }}(x)\right] \mathrm{%
Tr}\left[ \gamma _{5}\widetilde{S}_{b}^{e^{\prime }e}(-x)\gamma
_{5}S_{d}^{d^{\prime }d}(-x)\right] \right\} ,  \label{eq:CorrF2}
\end{eqnarray}%
where%
\begin{equation*}
\widetilde{S}_{b(q)}^{ij}(x)=CS_{b(q)}^{ij\mathrm{T}}(x)C.
\end{equation*}%
In expressions above $S_{q}^{ab}(x)$ and $S_{b}^{ab}(x)$ are the light $u,\
d\ $ and heavy $b$-quark propagators, respectively. We choose the light
quark propagator $S_{q}^{ab}(x)$ in the form%
\begin{eqnarray}
&&S_{q}^{ab}(x)=i\delta _{ab}\frac{\slashed x}{2\pi ^{2}x^{4}}-\delta _{ab}%
\frac{m_{q}}{4\pi ^{2}x^{2}}-\delta _{ab}\frac{\langle \overline{q}q\rangle
}{12}  \notag \\
&&+i\delta _{ab}\frac{\slashed xm_{q}\langle \overline{q}q\rangle }{48}%
-\delta _{ab}\frac{x^{2}}{192}\langle \overline{q}g_{\mathrm{s}}\sigma
Gq\rangle +i\delta _{ab}\frac{x^{2}\slashed xm_{q}}{1152}\langle \overline{q}%
g_{\mathrm{s}}\sigma Gq\rangle  \notag \\
&&-i\frac{g_{\mathrm{s}}G_{ab}^{\alpha \beta }}{32\pi ^{2}x^{2}}\left[ %
\slashed x{\sigma _{\alpha \beta }+\sigma _{\alpha \beta }}\slashed x\right]
-i\delta _{ab}\frac{x^{2}\slashed xg_{\mathrm{s}}^{2}\langle \overline{q}%
q\rangle ^{2}}{7776}  \notag \\
&&-\delta _{ab}\frac{x^{4}\langle \overline{q}q\rangle \langle g_{\mathrm{s}%
}^{2}G^{2}\rangle }{27648}+\ldots  \label{eq:qprop}
\end{eqnarray}%
For the $b$-quark propagator $S_{b}^{ab}(x)$ we employ the expression
\begin{eqnarray}
&&S_{b}^{ab}(x)=i\int \frac{d^{4}k}{(2\pi )^{4}}e^{-ikx}\Bigg \{\frac{\delta
_{ab}\left( {\slashed k}+m_{b}\right) }{k^{2}-m_{b}^{2}}  \notag \\
&&-\frac{g_{s}G_{ab}^{\alpha \beta }}{4}\frac{\sigma _{\alpha \beta }\left( {%
\slashed k}+m_{b}\right) +\left( {\slashed k}+m_{b}\right) \sigma _{\alpha
\beta }}{(k^{2}-m_{b}^{2})^{2}}  \notag \\
&&+\frac{g_{\mathrm{s}}^{2}G^{2}}{12}\delta _{ab}m_{b}\frac{k^{2}+m_{b}{%
\slashed k}}{(k^{2}-m_{b}^{2})^{4}}+\frac{g_{\mathrm{s}}^{3}G^{3}}{48}\delta
_{ab}\frac{\left( {\slashed k}+m_{b}\right) }{(k^{2}-m_{b}^{2})^{6}}  \notag
\\
&&\times \left[ {\slashed k}\left( k^{2}-3m_{b}^{2}\right) +2m_{b}\left(
2k^{2}-m_{b}^{2}\right) \right] \left( {\slashed k}+m_{b}\right) +\ldots %
\Bigg \}.  \notag \\
&&{}  \label{eq:Qprop}
\end{eqnarray}%
In Eqs.\ (\ref{eq:qprop}) and (\ref{eq:Qprop}) we use the notations
\begin{eqnarray}
&&G_{ab}^{\alpha \beta }=G_{A}^{\alpha \beta
}t_{ab}^{A},\,\,~~G^{2}=G_{\alpha \beta }^{A}G_{\alpha \beta }^{A},  \notag
\\
&&G^{3}=\,\,f^{ABC}G_{\mu \nu }^{A}G_{\nu \delta }^{B}G_{\delta \mu }^{C},
\label{eq:GG}
\end{eqnarray}%
where $A,B,C=1,\,2\,\ldots 8$. In Eq.\ (\ref{eq:GG}) $t^{A}=\lambda ^{A}/2$,
$\lambda ^{A}$ are the Gell-Mann matrices, and the gluon field strength
tensor $G_{\alpha \beta}^{A}\equiv G_{\alpha \beta }^{A}(0)$ is fixed at $%
x=0 $.

The QCD sum rule can be obtained by choosing the same Lorentz structures in
both of $\Pi _{\mu \nu }^{\mathrm{Phys}}(p)$ and $\Pi _{\mu \nu }^{\mathrm{%
QCD}}(p)$. We work with terms $\sim g_{\mu \nu }$, which do not contain
effects of spin-0 particles. The invariant amplitude $\Pi ^{\mathrm{QCD}%
}(p^{2})$ corresponding to this structure can be written down as the
dispersion integral
\begin{equation}
\Pi ^{\mathrm{QCD}}(p^{2})=\int_{4m_{b}^{2}}^{\infty }\frac{\rho ^{\mathrm{%
QCD}}(s)}{s-p^{2}}ds+...,  \label{CFQCD}
\end{equation}%
where $\rho ^{\mathrm{QCD}}(s)$ is the corresponding spectral density. It is
a key ingredient of sum rules for $m_{Z_{b}}^{2}$ and $f_{Z_{b}}^{2},$ and
can be obtained using the imaginary part of the invariant amplitude $\Pi ^{%
\mathrm{QCD}}(p^{2})$. Methods of such calculations are well known and
presented numerously in existing literature. Therefore, we omit further
details emphasizing only that $\rho^{\mathrm{QCD}}(s)$ in the present work
is calculated by including into analysis quark, gluon and mixed condensates
up to ten dimensions.

After applying the Borel transformation on the variable $p^{2}$ to $\Pi ^{%
\mathrm{QCD}}(p^{2})$, equating the obtained expression to $\mathcal{B}\Pi ^{%
\mathrm{Phys}}(p)$, and subtracting the continuum contribution, we obtain
the required sum rules. Thus, the mass of the $Z_{b}$ state can be evaluated
from the sum rule
\begin{equation}
m_{Z_{b}}^{2}=\frac{\int_{4m_{b}^{2}}^{s_{0}}dss\rho ^{\mathrm{QCD}%
}(s)e^{-s/M^{2}}}{\int_{4m_{b}^{2}}^{s_{0}}ds\rho ^{\mathrm{QCD}%
}(s)e^{-s/M^{2}}},  \label{eq:srmass}
\end{equation}%
whereas for the current coupling $\ f_{Z_{b}}$ we employ the formula
\begin{equation}
f_{Z_{b}}^{2}=\frac{1}{m_{Z_{b}}^{2}}\int_{4m_{b}^{2}}^{s_{0}}ds\rho ^{%
\mathrm{QCD}}(s)e^{(m_{Z_{b}}^{2}-s)/M^{2}}.  \label{eq:srcoupling}
\end{equation}%
The sum rules for $m_{Z_{b}}$ and $f_{Z_{b}}$ depend on different vacuum
condensates stemming from the quark propagators, on the mass of $b$-quark,
and on the Borel variable $M^{2}$ and continuum threshold $s_{0}$, which are
auxiliary parameters of numerical computations. The vacuum condensates are
parameters that do not depend on a problem under consideration: their
numerical values extracted once from some processes are applicable in all
sum rule computations. For quark and mixed condensates in the present work
we employ $\langle \bar{q}q\rangle =-(0.24\pm 0.01)^{3}~\mathrm{GeV}^{3}$, $%
\langle \overline{q}g_{\mathrm{s}}\sigma Gq\rangle =m_{0}^{2}\langle \bar{q}%
q\rangle $, where $m_{0}^{2}=(0.8\pm 0.1)~\mathrm{GeV}^{2}$, whereas for the
gluon condensates we utilize $\langle \alpha _{\mathrm{s}}G^{2}/\pi \rangle
=(0.012\pm 0.004)~\mathrm{GeV}^{4}$, $\langle g_{\mathrm{s}}^{3}G^{3}\rangle
=(0.57\pm 0.29)~\mathrm{GeV}^{6}$. The mass of the $b-$quark can be found in
Ref.\ \cite{Olive:2016xmw}: it is equal to $m_{b}=4.18_{-0.03}^{+0.04}~%
\mathrm{GeV}$.

The choice of the Borel parameter $M^2$ and continuum threshold $s_0$ should
obey some restrictions of sum rule calculations. Thus, limits within of
which $M^2$ can be varied (working window) are determined from convergence
of the operator product expansion and dominance of the pole contribution. In
the working window of the threshold parameter $s_0$ dependence of evaluating
quantities on $M^2$ should be minimal. In real calculations, however
quantities of interest depend on the parameters $M^2$ and $s_0$, which
affects an accuracy of extracted numerical values. Theoretical errors in sum
rule calculations may amount to $30 \%$ of obtained predictions, and
considerable part of these ambiguities are connected namely with a choice of
$M^2$ and $s_0$.

Analysis performed in accordance with these requirements allows us to fix
the working windows for $M^{2}$ and $s_{0}$:
\begin{equation}
M^{2}=9-12\ \mathrm{GeV}^{2},\ s_{0}=123-127\ \mathrm{GeV}^{2}.
\end{equation}%
In Figs.\ \ref{fig:Mass} and \ref{fig:Coup} we demonstrate results of
numerical computations of the mass $m_{Z_{b}}$ and current coupling $%
f_{Z_{b}}$ as functions of the parameters $M^{2}$ and $s_{0}$. As is seen, $%
m_{Z_{b}}$ and $f_{Z_{b}}$ are rather stable within working windows of the
auxiliary parameters, but there are still a dependence on them in plotted
figures. Our results for $m_{Z_{b}}$ and $f_{Z_{b}}$ read:
\begin{equation}
m_{Z_{b}}=10581_{-164}^{+142}\ \mathrm{MeV},\ \
f_{Z_{b}}=(2.79_{-0.65}^{+0.55})\cdot 10^{-2}\ \mathrm{GeV}^{4}.
\label{eq:ZbMF}
\end{equation}%
Within theoretical errors $m_{Z_{b}}$ is in agreement with experimental
measurements of the Belle Collaboration (\ref{eq:ExpMass}). The mass and
current coupling $Z_{b}$ given by Eq.\ (\ref{eq:ZbMF}) will be used as input
parameters in the next sections to find width of decays $Z_{b}\rightarrow
\Upsilon (nS)\pi $ and $Z_{b}\rightarrow h_{b}(mP)\pi $.

\begin{widetext}

\begin{figure}[h!]
\begin{center}
\includegraphics[totalheight=6cm,width=8cm]{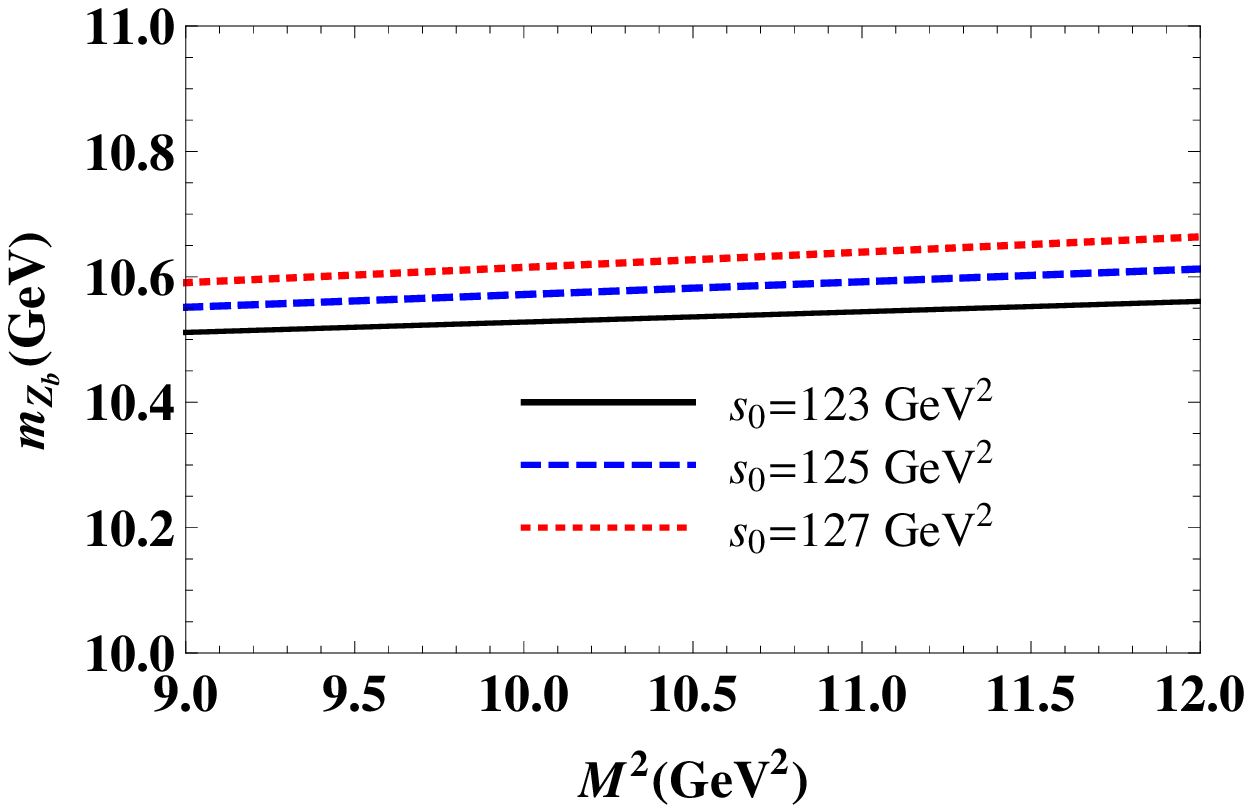}\,\,
\includegraphics[totalheight=6cm,width=8cm]{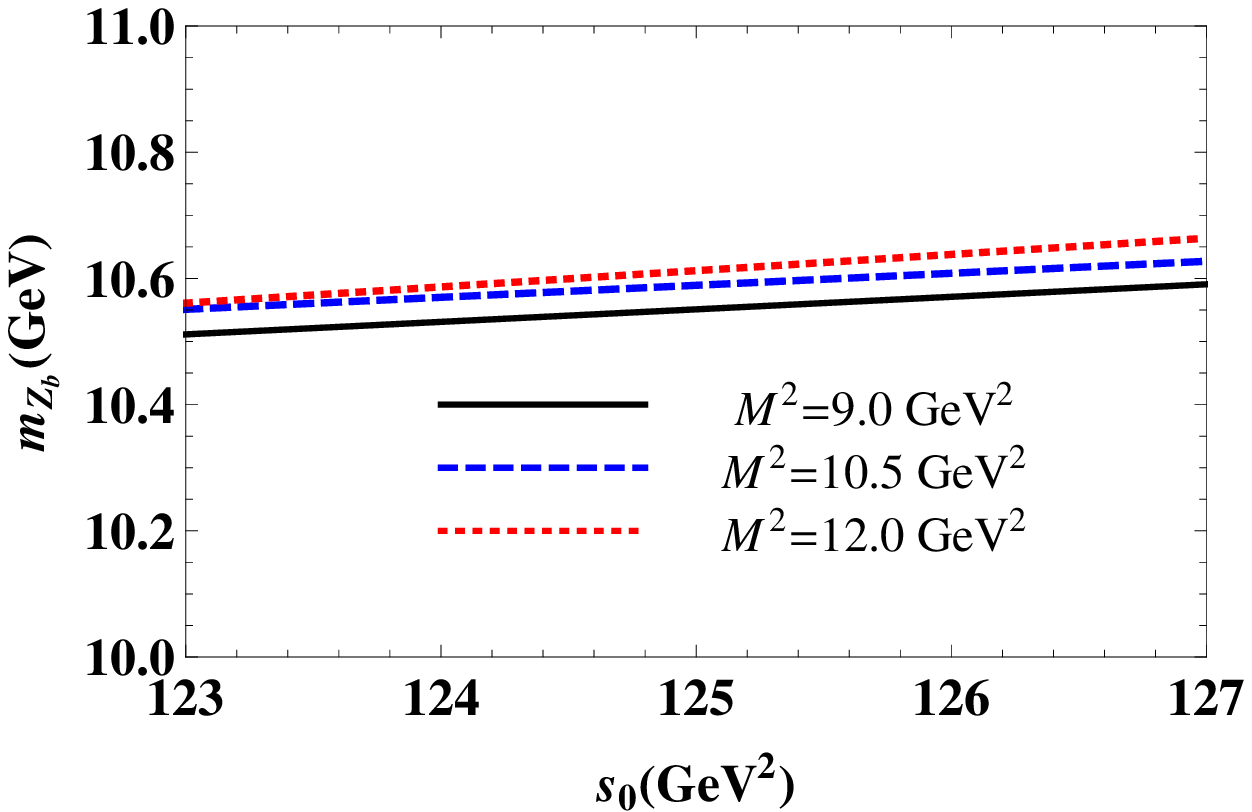}
\end{center}
\caption{ The mass of the $Z_b$ state vs  Borel parameter $M^2$ at fixed $s_0$
(left panel), and  continuum threshold $s_0$ at fixed $M^2$ (right panel).}
\label{fig:Mass}
\end{figure}

\begin{figure}[h!]
\begin{center}
\includegraphics[totalheight=6cm,width=8cm]{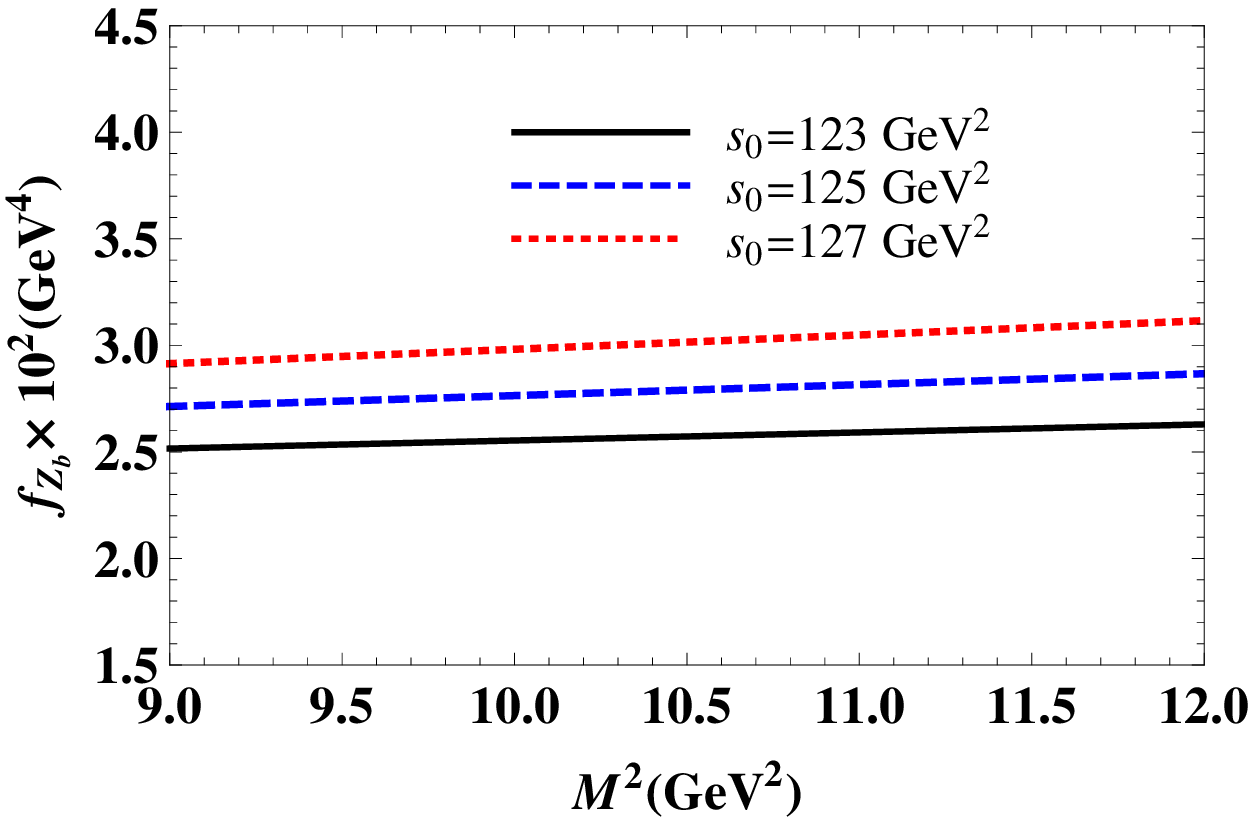}\,\,
\includegraphics[totalheight=6cm,width=8cm]{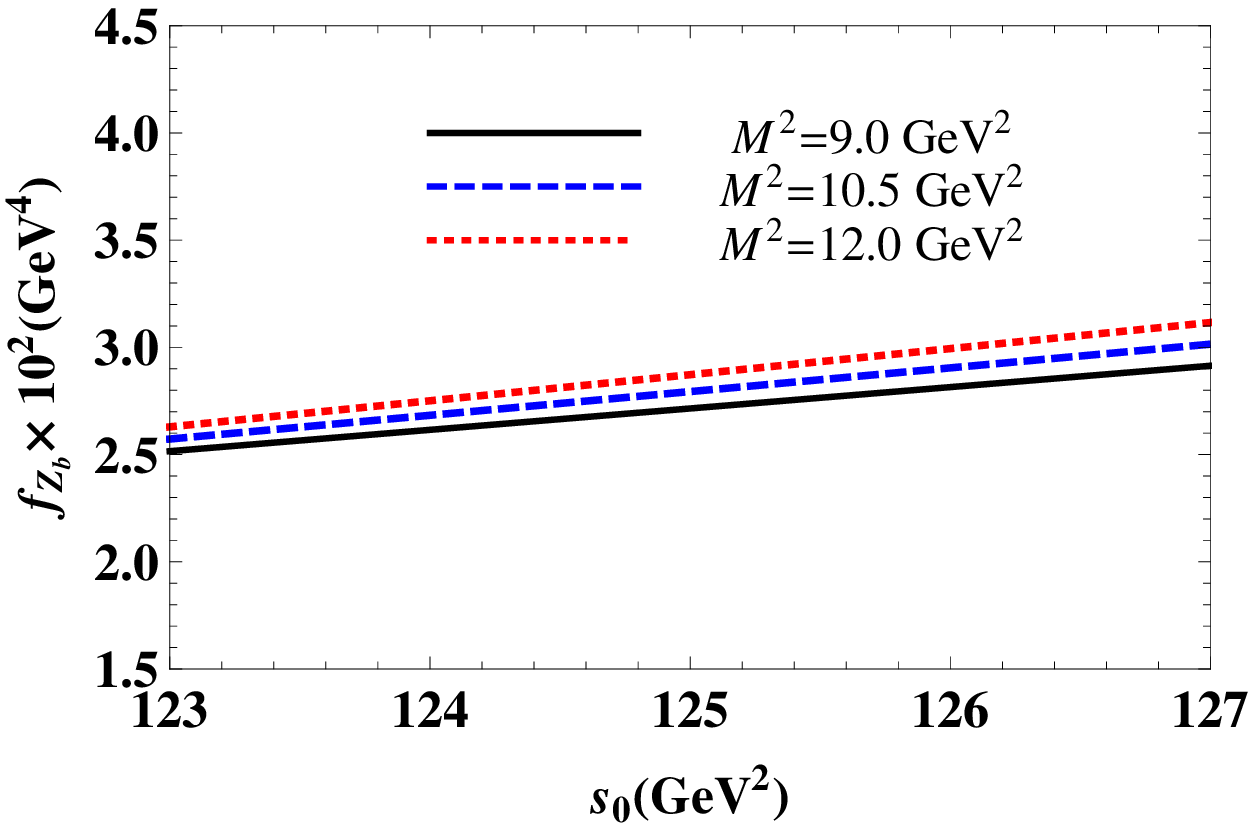}
\end{center}
\caption{ The dependence of the current coupling $f_{Z_b}$ of the $Z_b$ resonance
on the Borel parameter at chosen values of $s_0$ (left panel), and on the $s_0$ at
fixed $M^2$  (right panel).}
\label{fig:Coup}
\end{figure}

\end{widetext}

\section{Decay channels $Z_b \to \Upsilon(nS) \protect\pi, \ n=1,\ 2,\ 3. $}

\label{sec:Decays1}

This section is devoted to the calculation of the width of $Z_{b}\rightarrow
\Upsilon (nS)\pi ,\,\ n=1,\,2,\,3$ decays. To this end we determine the
strong couplings $g_{Z_{b}\Upsilon _{n}\pi },\ n=1,2,3$ (in formulas we
utilize $\Upsilon _{n}\equiv \Upsilon (nS)$ ) using QCD sum rules on the
light-cone in conjunction with ideas of a soft-meson approximation.

We start from analysis of the vertices $Z_{b}\Upsilon _{n}\pi $ aiming to
calculate $g_{Z_{b}\Upsilon _{n}\pi }$, and therefore consider the
correlation function
\begin{equation}
\Pi _{\mu \nu }(p,q)=i\int d^{4}xe^{ipx}\langle \pi (q)|\mathcal{T}\{J_{\mu
}^{\Upsilon }(x)J_{\nu }^{Z_{b}\dagger }(0)\}|0\rangle ,  \label{eq:CorrF3}
\end{equation}%
where
\begin{equation}
J_{\mu }^{\Upsilon }(x)=\overline{b}_{i}(x)\gamma _{\mu }b_{i}(x),
\end{equation}%
is the interpolating current for mesons $\Upsilon(nS)$. Here $p$, $q$ and $%
p^{\prime }=p+q$ are the momenta of $\Upsilon (nS)$, $\pi $ and $Z_{b}$,
respectively.

To derive sum rules for the couplings $g_{Z_{b}\Upsilon _{n}\pi }$, we
calculate $\Pi _{\mu \nu }(p,q)$ in terms of the physical degrees of
freedom. It is not difficult to obtain
\begin{eqnarray}
&&\Pi _{\mu \nu }^{\mathrm{Phys}}(p,q)=\sum_{n=1}^{3}\frac{\langle 0|J_{\mu
}^{\Upsilon }|\Upsilon _{n}\left( p\right) \rangle }{p^{2}-m_{\Upsilon
(nS)}^{2}}\langle \Upsilon _{n}\left( p\right) \pi (q)|Z_{b}(p^{\prime
})\rangle  \notag \\
&&\times \frac{\langle Z_{b}(p^{\prime })|J_{\nu }^{Z_{b}\dagger }|0\rangle
}{p^{\prime 2}-m_{Z_{b}}^{2}}+\ldots,  \label{eq:CorrF4}
\end{eqnarray}%
where the dots denote contribution of the higher resonances and continuum
states.

We introduce the matrix elements
\begin{eqnarray}
&&\langle 0|J_{\mu }^{\Upsilon }|\Upsilon _{n}\left( p\right) \rangle
=f_{\Upsilon _{n}}m_{\Upsilon _{n}}\varepsilon _{\mu },  \notag \\
&&\langle Z_{b}(p^{\prime })|J_{\nu }^{Z_{b}\dagger }|0\rangle
=f_{Z_{b}}m_{Z_{b}}\varepsilon _{\nu }^{\prime \ast },  \notag \\
&&\langle \Upsilon _{n}\left( p\right) \pi (q)|Z_{b}(p^{\prime })\rangle
=g_{Z_{b}\Upsilon _{n}\pi }\left[ (p\cdot p^{\prime })\right.  \notag \\
&&\left. \times (\varepsilon ^{\ast }\cdot \varepsilon ^{\prime })-(p\cdot
\varepsilon ^{\prime })(p^{\prime }\cdot \varepsilon ^{\ast })\right] ,
\label{eq:Mel}
\end{eqnarray}%
where $f_{\Upsilon _{n}},\,m_{\Upsilon _{n}},\,\varepsilon _{\mu }$ are the
decay constant, mass and polarization vector of the $\Upsilon (nS)$ meson,
and $\varepsilon _{\nu }^{\prime }$ is the polarization vector of the $Z_{b}$
state.

Having used Eq.\ (\ref{eq:Mel}) we rewrite the correlation function in the
form
\begin{eqnarray}
&&\Pi _{\mu \nu }^{\mathrm{Phys}}(p,q)=\sum_{n=1}^{3}\frac{g_{Z_{b}\Upsilon
_{n}\pi }f_{\Upsilon _{n}}f_{Z_{b}}m_{Z_{b}}m_{\Upsilon _{n}}}{\left(
p^{\prime 2}-m_{Z_{b}}^{2}\right) \left( p^{2}-m_{\Upsilon _{n}}^{2}\right) }
\notag \\
&&\times \left( \frac{m_{Z_{b}}^{2}+m_{\Upsilon _{n}}^{2}}{2}g_{\mu \nu
}-p_{\mu }^{\prime }p_{\nu }\right) +\ldots .  \label{eq:CorrF5}
\end{eqnarray}%
For calculation of the strong couplings we choose to work with the structure
$\sim g_{\mu \nu }$. To this end, we have to isolate the invariant function $%
\Pi ^{\mathrm{Phys}}(p^{2},p^{\prime 2})$ corresponding to this structure
and find its double Borel transformation. But, it is known that in the case
of vertices involving a tetraquark and two conventional mesons one has to
set $q=0$ \cite{Agaev:2016dev}. This is connected with the fact that
interpolating current for the tetraquark is composed of four quarks fields
and after contracting two of them in the correlation function $\Pi _{\mu \nu
}(p,q)$ with relevant quark fields from the heavy meson's current we
encounter a situation when remaining quarks are located at the same
space-time point. These quarks fields, sandwiched between a light meson and
vacuum instead of generating light meson's distribution amplitudes create
its local matrix elements. Then, in accordance with the four-momentum
conservation at such vertices we have to set $q=0$. In QCD light-cone sum
rules the limit $q\rightarrow 0$ when a light-cone expansion reduces to a
short-distant expansion over local matrix elements is known as a "soft-meson
approximation". The mathematical methods to handle soft-meson limit were
elaborated in Refs.\ \cite{Belyaev:1994zk,Ioffe:1983ju}, and were
successfully applied to tetraquark vertices in our works \cite%
{Agaev:2016ijz,Agaev:2016dsg,Agaev:2017foq,Agaev:2017tzv}. In soft limit $%
p^{\prime }=p$ and relevant invariant amplitudes in the correlation function
depend only on one variable $p^{2}$. In the present work we use this
approach which implies calculation of the correlation function with the
equal initial and final momenta $p^{\prime }=p$, and dealing with the
obtained double pole terms.

In fact, in the limit $p=p^{\prime }$ we replace in Eq.\ (\ref{eq:CorrF5})
\begin{equation*}
\frac{1}{\left( p^{\prime 2}-m_{Z_{b}}^{2}\right) \left( p^{2}-m_{\Upsilon
_{n}}^{2}\right) }
\end{equation*}%
by double pole factors
\begin{equation*}
\frac{1}{\left( p^{2}-m_{n}^{2}\right) ^{2}},
\end{equation*}%
where $m_{n}^{2}=(m_{Z_{b}}^{2}+m_{\Upsilon _{n}}^{2})/2$, and carry out the
Borel transformation over $p^{2}$. Then for the Borel transformation of $\Pi
^{\mathrm{Phys}}(p^{2})$ we get
\begin{eqnarray}
&&\mathcal{B}\Pi ^{\mathrm{Phys}}(p^{2})=\sum_{n=1}^{3}g_{Z_{b}\Upsilon
_{n}\pi }f_{\Upsilon _{n}}f_{Z_{b}}m_{Z_{b}}m_{\Upsilon _{n}}  \notag \\
&&\times m_{n}^{2}\frac{e^{-m_{n}^{2}/M^{2}}}{M^{2}}+\ldots .
\label{eq:CorrF5A}
\end{eqnarray}

Now one has to derive the correlation function in terms of the quark-gluon
degrees of freedom and find the QCD side of the sum rules. Contracting of
heavy quark fields in Eq.\ (\ref{eq:CorrF3}) yields
\begin{eqnarray}
&&\Pi _{\mu \nu }^{\mathrm{QCD}}(p,q)=\int d^{4}xe^{ipx}\frac{\epsilon
\widetilde{\epsilon }}{\sqrt{2}}\left[ \gamma _{5}\widetilde{S}%
_{b}^{ib}(x){}\gamma _{\mu }\right.  \notag \\
&&\left. \times \widetilde{S}_{b}^{ei}(-x){}\gamma _{\nu }+\gamma _{\nu }%
\widetilde{S}_{b}^{ib}(x){}\gamma _{\mu }\widetilde{S}_{b}^{ei}(-x){}\gamma
_{5}\right] _{\alpha \beta }  \notag \\
&&\times \langle \pi (q)|\overline{u}_{\alpha }^{a}(0)d_{\beta
}^{d}(0)|0\rangle ,  \label{eq:CorrF6}
\end{eqnarray}%
where $\alpha $ and $\beta $ are the spinor indices. We continue and use the
expansion
\begin{equation}
\overline{u}_{\alpha }^{a}d_{\beta }^{d}\rightarrow \frac{1}{4}\Gamma
_{\beta \alpha }^{j}\left( \overline{u}^{a}\Gamma ^{j}d^{d}\right) ,
\label{eq:MatEx}
\end{equation}%
where $\Gamma ^{j}$ is the full set of Dirac matrixes
\begin{equation*}
\Gamma ^{j}=\mathbf{1,\ }\gamma _{5},\ \gamma _{\lambda },\ i\gamma
_{5}\gamma _{\lambda },\ \sigma _{\lambda \rho }/\sqrt{2}.
\end{equation*}%
Replacing $\overline{u}_{\alpha }^{a}d_{\beta }^{d}$ in Eq.\ (\ref{eq:CorrF6}%
) by this expansion and performing summations over color indices it is not
difficult to determine local matrix elements of pion which contribute to $%
\Pi _{\mu \nu }^{\mathrm{QCD}}(p,q)$ (see, Ref.\ \cite{Agaev:2016dev} for
details). It turns out that in soft limit the pion's local matrix element
which contributes to $\mathrm{Im}\Pi _{\mu \nu }^{\mathrm{QCD}}(p,q=0)$ is
\begin{equation}
\langle 0|\overline{d}(0)i\gamma _{5}u(0)|\pi (q)\rangle =f_{\pi }\mu _{\pi
},  \label{eq:MatE2}
\end{equation}%
where
\begin{equation*}
\mu _{\pi }=\frac{m_{\pi }^{2}}{m_{u}+m_{d}}.
\end{equation*}%
After fixing in $\mathrm{Im}\Pi _{\mu \nu }^{\mathrm{QCD}}(p,q=0)$ the
structure $\sim g_{\mu \nu }$ it is straightforward to extract $\rho
_{\Upsilon }^{\mathrm{QCD}}(s)$ as a sum of the perturbative and
nonperturbative components:
\begin{equation}
\rho _{\Upsilon }^{\mathrm{QCD}}(s)=\frac{f_{\pi }\mu _{\pi }}{12\sqrt{2}}%
\left[ \rho ^{\mathrm{pert.}}(s)+\rho ^{\mathrm{n.-pert.}}(s)\right]
\label{eq:SDensity}
\end{equation}%
The $\rho _{\Upsilon }^{\mathrm{QCD}}(s)$ can be obtained after replacement $%
m_{c}\rightarrow m_{b}$ from the spectral density of $\ Z_{c}\rightarrow
J/\psi \pi $ decay calculated in Refs.\ \cite{Agaev:2016dev,Agaev:2017tzv}.
Its perturbative component $\rho ^{\mathrm{pert.}}(s)$ has a simple form and
reads%
\begin{equation}
\rho ^{\mathrm{pert.}}(s)=\frac{(s+2m_{b}^{2})\sqrt{s(s-4m_{b}^{2})}}{\pi
^{2}s}.
\end{equation}%
The nonperturbative contribution $\rho ^{\mathrm{n.-pert.}}(s)$ depends on
the vacuum expectation values of the gluon operators and contains terms of
four, six and eight dimensions. Its explicit expression was presented in
Appendix of Ref.\ \cite{Agaev:2017tzv}.

The continuum subtraction in the case under consideration can be done using
the quark-hadron duality, which lead the desired sum rule for strong
couplings. We get:
\begin{eqnarray}
&&\sum_{n=1}^{3}g_{Z_{b}\Upsilon _{n}\pi }f_{\Upsilon
_{n}}f_{Z_{b}}m_{Z_{b}}m_{\Upsilon _{n}}m_{n}^{2}\frac{e^{-m_{n}^{2}/M^{2}}}{%
M^{2}}  \notag \\
&=&\int_{4m_{b}^{2}}^{s_{0}}dse^{-s/M^{2}}\rho _{\Upsilon }^{\mathrm{QCD}%
}(s).  \label{eq:DecaySR}
\end{eqnarray}%
Here some comments are in order on obtained expression (\ref{eq:DecaySR}).
It is known, that the soft limit considerably simplifies the QCD side of
light-cone sum rule expressions \cite{Belyaev:1994zk}. At the same time, in
the limit $q\rightarrow 0$ the phenomenological side of the sum rules gains
contributions which are not suppressed relative to a main term. In our case
the main term corresponds to vertex $Z_{b}\Upsilon (1S)\pi $, where the
tetraquark and mesons are ground-state particles. Additional contributions
emerge due to vertices $Z_{b}\Upsilon \pi $ where some of particles (or all
of them) are on their excited states. In Eq.\ (\ref{eq:DecaySR}) terms
corresponding to vertices $Z_{b}\Upsilon (2S)\pi $ and $Z_{b}\Upsilon
(3S)\pi $ belong to this class of contributions. When we are interested in
extraction of parameters of a vertex built of only ground-state particles
these additional contributions are undesired contaminations which may affect
accuracy of calculations. A technique to eliminate them from sum rules is
also well known \cite{Belyaev:1994zk,Ioffe:1983ju}. To this end, in
accordance with elaborated recipies one has to act by the operator
\begin{equation}
\mathcal{P}(M^{2},m_{n}^{2})=\left( 1-M^{2}\frac{d}{dM^{2}}\right)
M^{2}e^{m_{n}^{2}/M^{2}},  \label{eq:softop}
\end{equation}%
to Eq.\ (\ref{eq:DecaySR}). In the present work we are going to evaluate
three strong couplings $g_{Z_{b}\Upsilon _{n}\pi }$ and, therefore use the
original form of the sum rule given by Eq.\ (\ref{eq:DecaySR}). But it
provides only one equality for three unknown quantities. In order to get two
additional equations we act by operators $d/d(-1/M^{2})$ and $%
d^{2}/d(-1/M^{2})^{2}$ to both sides of Eq.\ (\ref{eq:DecaySR}) and solve
obtained equations to find $g_{Z_{b}\Upsilon _{n}\pi }$.

The width of the decays $Z_{b}\rightarrow \Upsilon (nS)\pi ,\ n=1,2,3$ can
be calculated applying the standard methods and has the same form as in the
case of the decay $Z_{c}\rightarrow J/\psi \pi $. After evident replacements
in corresponding formula we get:
\begin{eqnarray}
&&\Gamma \left( Z_{b}\rightarrow \Upsilon _{n}\pi \right) =\frac{%
g_{Z_{b}\Upsilon _{n}\pi }^{2}m_{\Upsilon _{n}}^{2}}{24\pi }\lambda \left(
m_{Z_{b}},\ m_{\Upsilon _{n}},m_{\pi }\right)  \notag \\
&&\times \left[ 3+\frac{2\lambda ^{2}\left( m_{Z_{b}},\ m_{\Upsilon
_{n}},m_{\pi }\right) }{m_{\Upsilon _{n}}^{2}}\right] ,  \label{eq:DW}
\end{eqnarray}%
where
\begin{equation*}
\lambda (a,\ b,\ c)=\frac{\sqrt{a^{4}+b^{4}+c^{4}-2\left(
a^{2}b^{2}+a^{2}c^{2}+b^{2}c^{2}\right) }}{2a}.
\end{equation*}

The key component in Eq.\ (\ref{eq:DW}) is the strong coupling $%
g_{Z_{b}\Upsilon _{n}\pi }$. Relevant sum rules contain spectroscopic
parameters of the tetraquark $Z_{b}$, and mesons $\Upsilon (nS)$ and $\pi $.
The mass and current coupling of the $Z_{b}$ resonance have been calculated
in the previous section. For numerical computations we take masses $%
m_{\Upsilon _{n}}$ and decay constants $f_{\Upsilon _{n}}$ of the mesons $%
\Upsilon (nS)$ from Ref.\ \cite{Olive:2016xmw}. The relevant information is
shown in Table\ \ref{tab:Param}.
\begin{table}[tbp]
\begin{tabular}{|c|c|}
\hline\hline
Parameters & Values (in ($\mathrm{MeV}$) \\ \hline\hline
$m_{\Upsilon_{1}}$ & $9460.30 \pm 0.26$ \\
$f_{\Upsilon_{1}}$ & $708 \pm 8 $ \\
$m_{\Upsilon_{2}}$ & $10023.26 \pm 0.31 $ \\
$f_{\Upsilon_{2}}$ & $482 \pm 10 $ \\
$m_{\Upsilon_{3}}$ & $10355.2 \pm 0.5$ \\
$f_{\Upsilon_{3}}$ & $346 \pm 50$ \\
$m_{\pi}$ & $139.57061 \pm 0.00024$ \\
$f_{\pi}$ & $131.5 $ \\ \hline\hline
\end{tabular}%
\caption{Spectroscopic parameters of the mesons $\Upsilon _{nS}$ and $%
\protect\pi $.}
\label{tab:Param}
\end{table}

In calculations the Borel parameter $M^{2}$ and continuum threshold $s_{0}$
are varied within regions
\begin{equation}
M^{2}=10-13\ \mathrm{GeV}^{2},\ s_{0}=124-128\ \mathrm{GeV}^{2},
\end{equation}%
which are almost identical to similar working windows in the mass and
current coupling calculations being slightly shifted towards larger values.

For the couplings $g_{Z_{b}\Upsilon _{n}\pi }$ we obtain (in $\mathrm{GeV}%
^{-1}$):
\begin{eqnarray}
&&g_{Z_{b}\Upsilon _{1}\pi }=0.019\pm 0.005,\ \,g_{Z_{b}\Upsilon _{2}\pi
}=0.090\pm 0.031,  \notag \\
&&g_{Z_{b}\Upsilon _{3}\pi }=0.104\pm 0.031.
\end{eqnarray}%
For the width of the decays $Z_{b}\rightarrow \Upsilon (nS)\pi $ these
couplings lead to predictions
\begin{eqnarray}
&&\Gamma (Z_{b}\rightarrow \Upsilon (1S)\pi )=1.36\pm 0.43\ \mathrm{MeV},
\notag \\
&&\Gamma (Z_{b}\rightarrow \Upsilon (2S)\pi )=17.18\pm 5.01\ \mathrm{MeV},
\notag \\
&&\Gamma (Z_{b}\rightarrow \Upsilon (3S)\pi )=8.27\pm 2.69\ \mathrm{MeV}.
\label{eq:ResultsDW1}
\end{eqnarray}%
Obtained predictions for width of the decays $\Gamma (Z_{b}\rightarrow
\Upsilon (nS)\pi )$ are final results of this section and will be used for
comparison with the experimental data.

\section{$Z_b \to h_b(1P) \protect\pi$ and $Z_b \to h_b(2P) \protect\pi $
decays}

\label{sec:Decays2}
The second class of decays which we consider contains two processes $%
Z_{b}\rightarrow h_{b}(mP)\pi ,\ m=1,2$. We follow the same prescriptions as
in the case of $Z_{b}\rightarrow \Upsilon (nP)\pi $ decays and derive sum
rules for the strong couplings $g_{Z_{b}h_{b}\pi }$ and $g_{Z_{b}h_{b}^{%
\prime }\pi }$ (hereafter we employ short-hand notations $h_{b}\equiv
h_{b}(1P)$ and $h_{b}^{\prime }\equiv h_{b}(2P)$). From analysis performed
in the previous section it is clear that corresponding sum rules will depend
on numerous input parameters including mass and decay constant of the mesons
$h_{b}(1P)$ and $h_{b}(2P)$. Information on the spectroscopic parameters of $%
h_{b}(1P)$ is available in the literature. Indeed, in the context of QCD sum
rule method mass and decay constant of $h(1P)$ were calculated in Ref.\ \cite%
{Wang:2012gj}. But decay constant $f_{h_{b}^{\prime }}$ of the meson $%
h_{b}(2P)$ was not evaluated, therefore in the present work we have first to
find the parameters $m_{h_{b}^{\prime }}$ and $f_{h_{b}^{\prime }}$, and
turn after that to our main task.


\subsection{Spectroscopic parameters of the mesons $h_{b}(1P)$ and $%
h_{b}(2P) $}

The meson $h(1P)$ is the spin-singlet $P$-wave bottomonium with quantum
numbers $J^{PC}=1^{+-}$, whereas $h(2P)$ is its first radial excitation.
Parameters of the $h_{b}(1P)$ and $h_{b}(2P)$ mesons in the framework of QCD
two-point sum rule method can be extracted from the correlation function
\begin{equation}
\Pi _{\mu \nu \alpha \beta }(p)=i\int d^{4}xe^{ipx}\langle 0|\mathcal{T}%
\left\{ J_{\mu \nu }^{h}(x)J_{\alpha \beta }^{h\dagger }(0)\right\}
|0\rangle ,
\end{equation}%
where the interpolating current for $h_{b}(mP)$ mesons is chosen as
\begin{equation}
J_{\mu \nu }^{h}(x)=\overline{b}^{i}(x)\sigma _{\mu \nu }\gamma _{5}b^{i}(x).
\end{equation}%
It couples both to $h_{b}(1P)$ and $h_{b}(2P)$, and is convenient for
analysis of $J^{PC}=1^{+-}$ mesons (see, Ref.\ \cite{Wang:2012gj}).

In order to find required sum rules we use "ground-state+first radial
excitation+continuum" scheme. Then, the physical side of the sum rule
\begin{eqnarray}
&&\Pi _{\mu \nu \alpha \beta }^{\mathrm{Phys}}(p)=\frac{\langle 0|J_{\mu \nu
}^{h}|h_{b}(p)\rangle \langle h_{b}(p)|J_{\alpha \beta }^{h\dagger
}(0)|0\rangle }{m_{h_{b}}^{2}-p^{2}}  \notag \\
&&+\frac{\langle 0|J_{\mu \nu }^{h}|h_{b}^{\prime }(p)\rangle \langle
h_{b}^{\prime }(p)|J_{\alpha \beta }^{h\dagger }(0)|0\rangle }{%
m_{h_{b}^{\prime }}^{2}-p^{2}}+\ldots ,
\end{eqnarray}%
contains two terms of interest and also contribution of higher resonances
and continuum states denoted by dots. We continue by introducing the matrix
elements
\begin{equation}
\langle 0|J_{\mu \nu }^{h}|h_{b}^{(\prime )}(p)\rangle =f_{h_{b}^{(\prime
)}}(\varepsilon _{\mu }^{(\prime )}p_{\nu }-\varepsilon _{\nu }^{(\prime
)}p_{\mu }),
\end{equation}%
and recast the correlation function $\Pi _{\mu \nu \alpha \beta }^{\mathrm{%
Phys}}(p)$ into the form
\begin{eqnarray}
&&\Pi _{\mu \nu \alpha \beta }^{\mathrm{Phys}}(p)=\frac{f_{h_{b}}^{2}}{%
m_{h_{b}}^{2}-p^{2}}\left[ \widetilde{g}_{\mu \alpha }p_{\nu }p_{\beta }-%
\widetilde{g}_{\mu \beta }p_{\nu }p_{\alpha }\right.  \notag \\
&&\left. -\widetilde{g}_{\nu \alpha }p_{\mu }p_{\beta }+\widetilde{g}_{\nu
\beta }p_{\mu }p_{\alpha }\right] +\frac{f_{h_{b}^{\prime }}^{2}}{%
m_{h_{b}^{\prime }}^{2}-p^{2}}\left[ \widetilde{g}_{\mu \alpha }^{\prime
}p_{\nu }p_{\beta }\right.  \notag \\
&&\left. -\widetilde{g}_{\mu \beta }^{\prime }p_{\nu }p_{\alpha }-\widetilde{%
g}_{\nu \alpha }^{\prime }p_{\mu }p_{\beta }+\widetilde{g}_{\nu \beta
}^{\prime }p_{\mu }p_{\alpha }\right] ,  \label{eq:CorrF7}
\end{eqnarray}%
where
\begin{equation*}
\widetilde{g}_{\mu \alpha }^{(\prime )}=-g_{\mu \alpha }+\frac{p_{\mu
}p_{\alpha }}{m_{h_{b}^{(\prime )}}^{2}}.
\end{equation*}%
The Borel transformation of $\Pi _{\mu \nu \alpha \beta }^{\mathrm{Phys}}(p)$
can be obtained by simple replacements in Eq.\ (\ref{eq:CorrF7})
\begin{equation*}
\mathcal{B}\frac{f_{h_{b}^{(\prime )}}^{2}}{m_{h_{b}^{(\prime )}}^{2}-p^{2}}%
=f_{h_{b}^{(\prime )}}^{2}e^{-m_{h_{b}^{(\prime )}}^{2}/M^{2}}.
\end{equation*}%
The obtained by this way expression contains numerous Lorentz structures
which, in general, may be employed to derive sum rules for masses and decay
constants: We choose a structure $\sim \widetilde{g}_{\mu \alpha }p_{\nu
}p_{\beta }$ to extract sum rules. The term with the same structure should
be isolated in the Borel transformation of $\Pi _{\mu \nu \alpha \beta }^{%
\mathrm{QCD}}(p)$, i. e. in expression of the correlation function
calculated using quark-gluon degrees of freedom.

After simple computations for $\Pi _{\mu \nu \alpha \beta }^{\mathrm{QCD}%
}(p) $ we get
\begin{eqnarray}
\Pi _{\mu \nu \alpha \beta }^{\mathrm{QCD}}(p) &=&i\int d^{4}xe^{ipx}\mathrm{%
Tr}\left[ \gamma _{5}\sigma _{\alpha \beta }S_{b}^{ji}(-x)\right.  \notag \\
&&\left. \times \sigma _{\mu \nu }\gamma _{5}S_{b}^{ij}(x)\right] .
\end{eqnarray}%
The following operations are standard manipulations; they imply Borel
transforming of $\Pi _{\mu \nu \alpha \beta }^{\mathrm{QCD}}(p)$, equating
the structures $\sim \widetilde{g}_{\mu \alpha }p_{\nu }p_{\beta }$ in both
the physical and QCD sides of obtained equality, and subtracting the
continuum contribution. We obtain the second sum rule by acting on first one
by $d/d(-1/M^{2}) $. These two sum rules allow us to evaluate masses and
decay constants of the $h_{b}(1P)$ and $h_{b}(2P)$ mesons. At the first
stage we employ "ground-state +continuum" scheme, which is commonly used in
sum rule computations. This means that we include the excited $h_{b}(2P)$
meson into the "higher resonances and continuum" part of sum rules and fix
working windows for $M^{2}$ and $s_{0}$. From these sum rules we extract
spectroscopic parameters of the $h_{b}(1P)$ meson $m_{h_{b}}$ and $f_{h_{b}}$%
. At the next step we employ the same sum rules with $s_{0}^{\ast }>s_{0}$
to embrace contribution arising from $h_{b}(2P)$, and treat $m_{h_{b}}$ and $%
f_{h_{b}}$ evaluated at the first stage as fixed parameters.

Numerical analysis restricts variation of the parameters $M^{2}$ and $s_{0}$
within the regions
\begin{equation*}
M^{2}=10-12\ \mathrm{GeV}^{2},\ s_{0}=103-105\ \mathrm{GeV}^{2},
\end{equation*}%
and we find%
\begin{equation}
m_{h_{b}}=9886_{-78}^{+81}\ \mathrm{MeV,\ \ }f_{h_{b}}=325_{-57}^{+61}\ \
\mathrm{MeV}.  \label{eq:H1mass}
\end{equation}%
At the next step we use
\begin{equation*}
\ s_{0}^{\ast }=109-111\ \mathrm{GeV}^{2},
\end{equation*}%
and get%
\begin{equation}
m_{h_{b}^{\prime }}=10331_{-117}^{+108}\ \mathrm{MeV,\ \ }f_{h_{b}^{\prime
}}=286_{-53}^{+58}\ \ \mathrm{MeV}.  \label{eq:H2mass}
\end{equation}%
Parameters of the $h_{b}(2P)$ meson are among essentially new results of the
present work, therefore in Figs.\ \ref{fig:Masshb2} and \ref{fig:Couphb2} we
demonstrate $m_{h_{b}(2P)}$ and $f_{h_{b}(2P)}$ as functions of the Borel
parameter $M^{2}$ and continuum threshold $s_{0}$.

Comparing our results with experimental information on masses of the $%
h_{b}(mP)$ mesons \cite{Olive:2016xmw}
\begin{eqnarray*}
m_{h_{b}} &=&9899.3\pm 0.8\ \mathrm{MeV},\  \\
m_{h_{b}^{\prime }} &=&10259.8\pm 1.2\ \mathrm{MeV},
\end{eqnarray*}%
we see a reasonable agreement between them.

\begin{widetext}

\begin{figure}[h!]
\begin{center}
\includegraphics[totalheight=6cm,width=8cm]{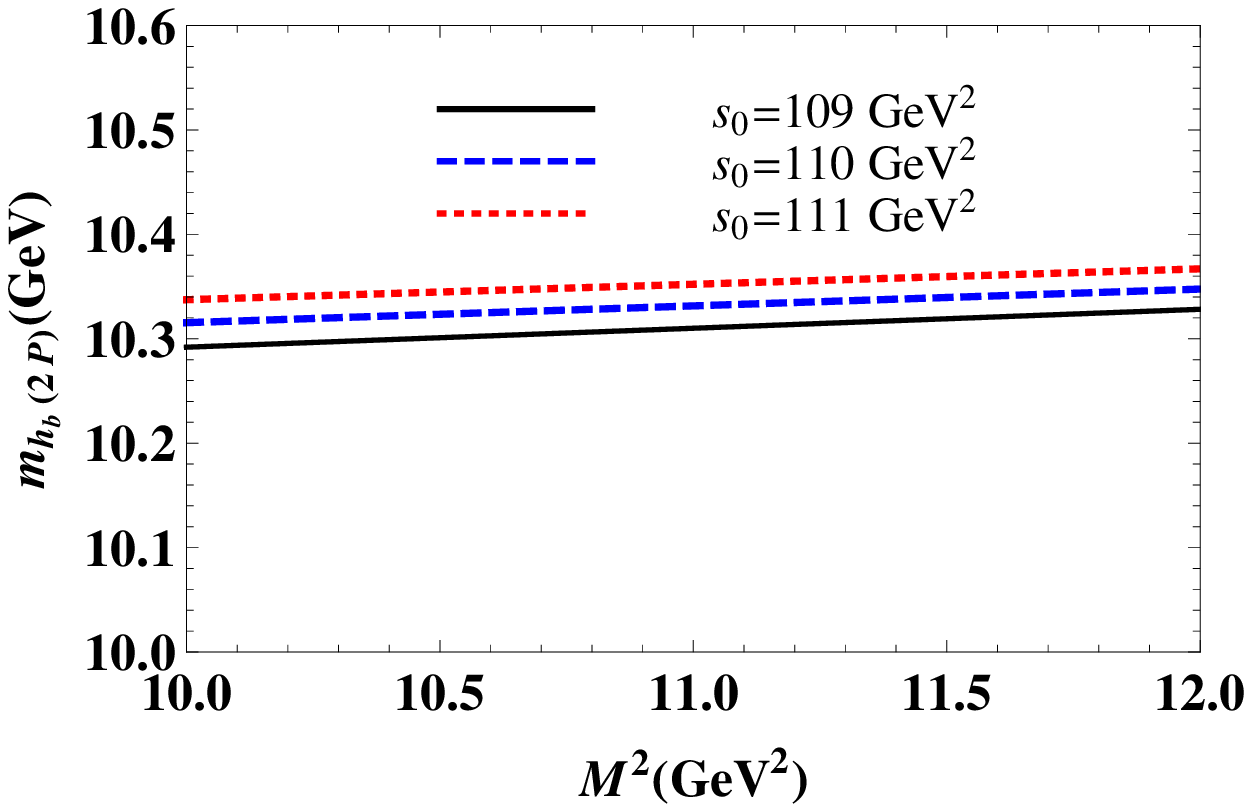}\,\,
\includegraphics[totalheight=6cm,width=8cm]{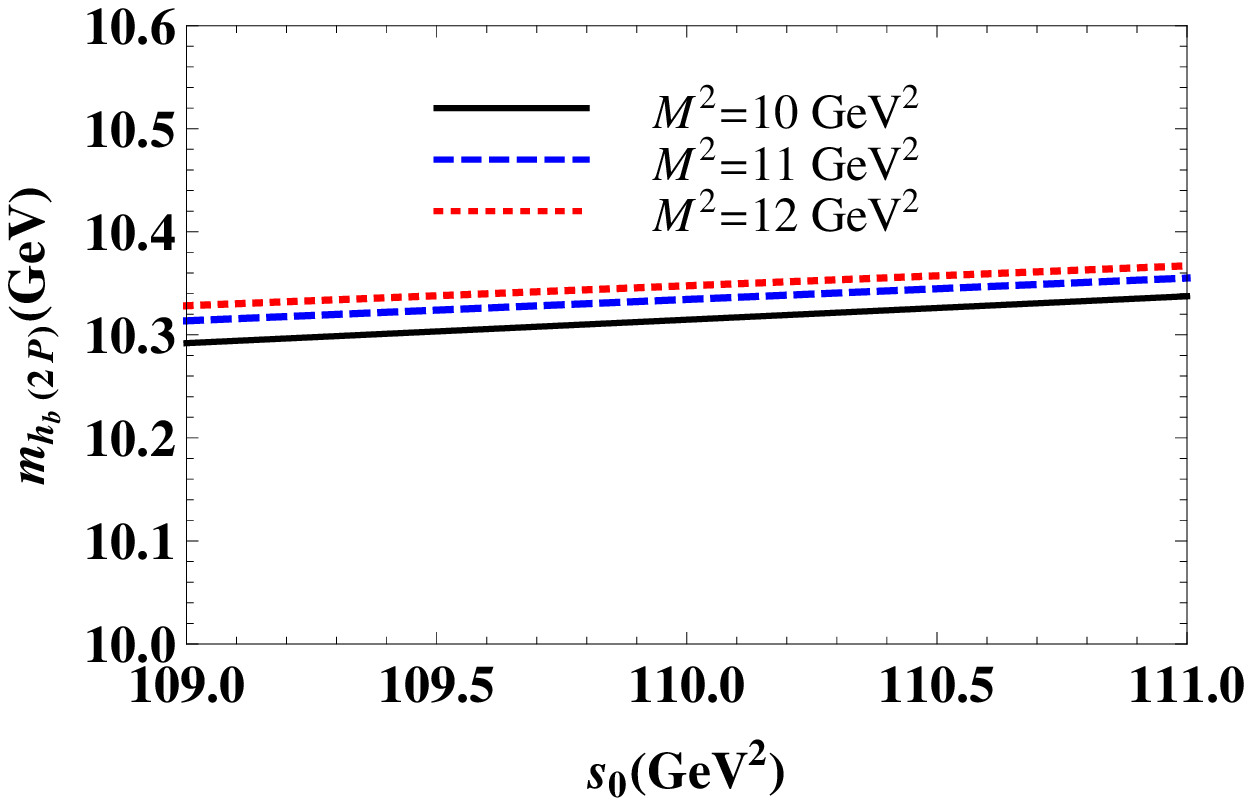}
\end{center}
\caption{ The mass of the meson $h_b(2P)$  as a function of the Borel parameter
$M^2$ at fixed $s_0$ (left panel), and as a function of the continuum threshold
$s_0$ at fixed $M^2$ (right panel).}
\label{fig:Masshb2}
\end{figure}

\begin{figure}[h!]
\begin{center}
\includegraphics[totalheight=6cm,width=8cm]{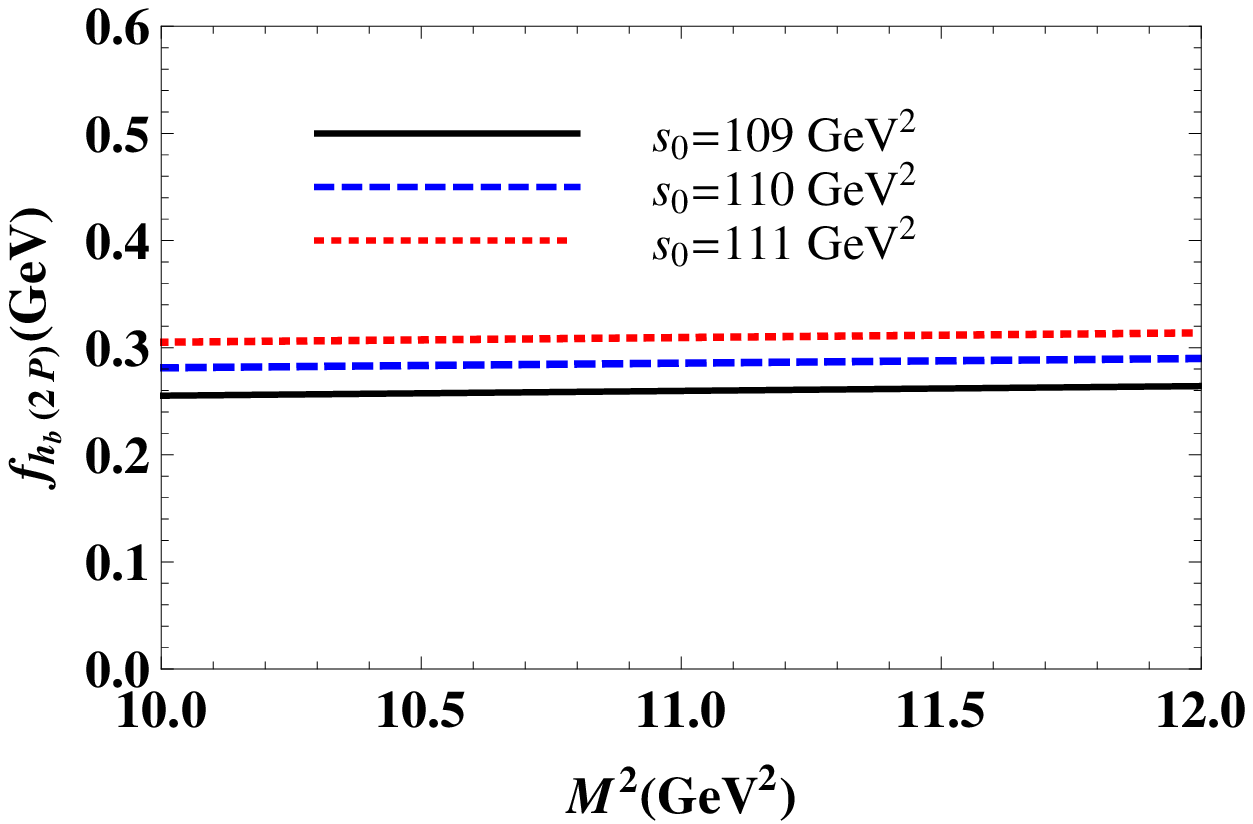}\,\,
\includegraphics[totalheight=6cm,width=8cm]{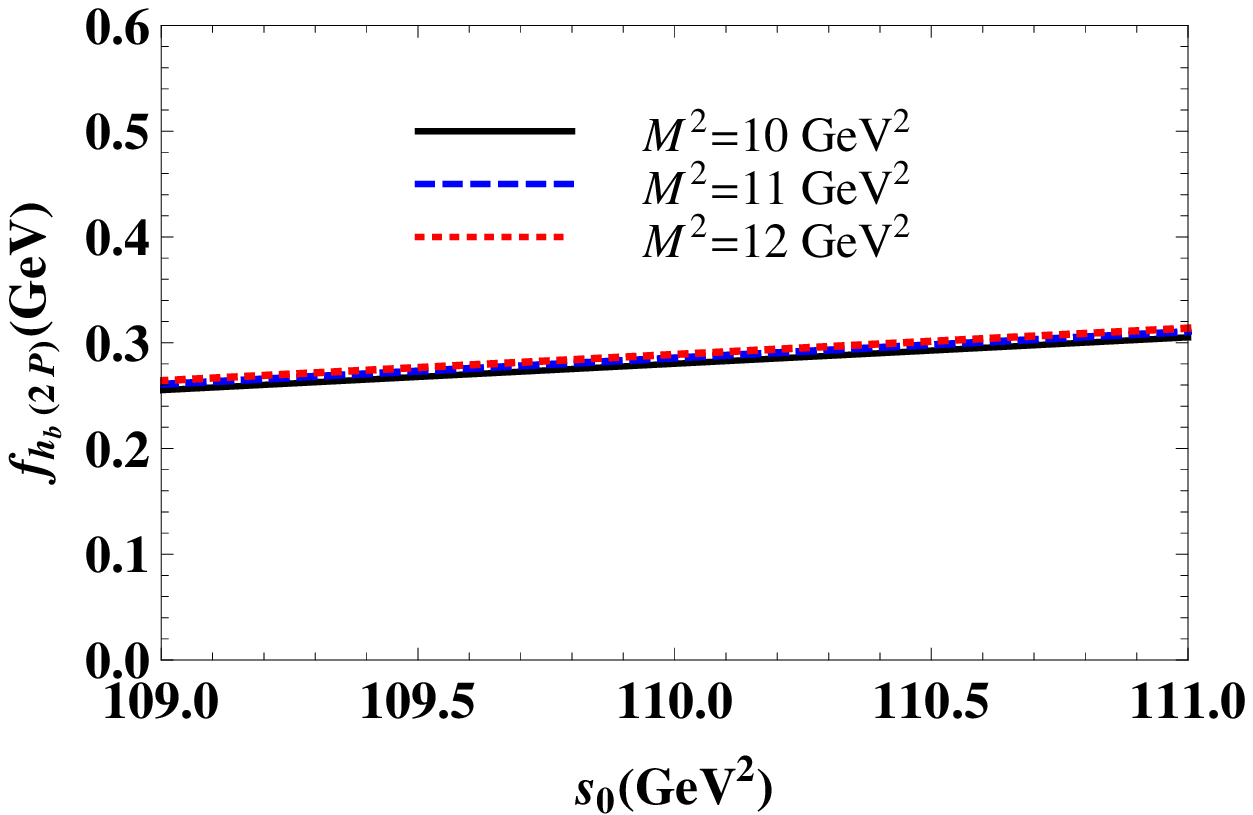}
\end{center}
\caption{ The dependence of the decay constant $f_{h_b(2P)}$  on the
Borel parameter at chosen values of $s_0$ (left panel), and on the $s_0$
at fixed $M^2$  (right panel).}
\label{fig:Couphb2}
\end{figure}

\end{widetext}


\subsection{Width of decays $Z_{b}\rightarrow h_{b}(1P)\protect\pi $ and $%
Z_{b}\rightarrow h_{b}(2P)\protect\pi $}

Analysis of the vertices $Z_{b}h_{b}(mP)\pi $ does not differ from analogous
investigation carried out in the previous section. We start here from the
correlator
\begin{equation*}
\Pi _{\mu \nu \lambda }(p,q)=i\int d^{4}xe^{ipx}\langle \pi (q)|\mathcal{T}%
\{J_{\mu \nu }^{h}(x)J_{\lambda }^{Z_{b}\dagger }(0)\}|0\rangle ,
\end{equation*}%
and for its phenomenological representation get%
\begin{eqnarray}
&&\Pi _{\mu \nu \lambda }^{\mathrm{Phys}}(p,q)=\frac{\langle 0|J_{\mu \nu
}^{h}|h_{b}\left( p\right) \rangle }{p^{2}-m_{h_{b}}^{2}}\langle h_{b}\left(
p\right) \pi (q)|Z_{b}(p^{\prime })\rangle  \notag \\
&&\times \frac{\langle Z_{b}(p^{\prime })|J_{\lambda }^{Z_{b}\dagger
}|0\rangle }{p^{\prime 2}-m_{Z_{b}}^{2}}+\frac{\langle 0|J_{\mu \nu
}^{h}|h_{b}^{\prime }\left( p\right) \rangle }{p^{2}-m_{h_{b}^{\prime }}^{2}}
\\
&&\times \langle h_{b}^{\prime }\left( p\right) \pi (q)|Z_{b}(p^{\prime
})\rangle \frac{\langle Z_{b}(p^{\prime })|J_{\lambda }^{Z_{b}\dagger
}|0\rangle }{p^{\prime 2}-m_{Z_{b}}^{2}}\ldots .
\end{eqnarray}%
The $\Pi _{\mu \nu \lambda }^{\mathrm{Phys}}(p,q)$ contains two terms of
initerest and contributions coming from higher resonances and continuum
shown above as dots. Using matrix elements of the currents $J_{\mu \nu }^{h}$
and $J_{\lambda }^{Z_{b}}$ and introducing the vertex%
\begin{equation}
\langle h_{b}^{(\prime )}\left( p\right) \pi (q)|Z_{b}(p^{\prime })\rangle
=g_{Z_{b}h_{b}^{(\prime )}\pi }\epsilon _{\alpha \beta \gamma \delta
}\varepsilon _{\alpha }^{\ast }(p)\varepsilon _{\beta }^{\prime }(p^{\prime
})p_{\gamma }p_{\delta }^{\prime },
\end{equation}%
we find%
\begin{eqnarray}
&&\Pi _{\mu \nu \lambda }^{\mathrm{Phys}}(p,q)=\frac{f_{Z_{b}}m_{Z_{b}}}{%
\left( p^{\prime 2}-m_{Z_{b}}^{2}\right) }\left[ \frac{g_{Z_{b}h_{b}\pi
}f_{h_{b}}}{\left( p^{2}-m_{h_{b}}^{2}\right) }+\frac{g_{Z_{b}h_{b}^{\prime
}\pi }f_{h_{b}^{\prime }}}{\left( p^{2}-m_{h_{b}^{\prime }}^{2}\right) }%
\right]  \notag \\
&&\times \left( \epsilon _{\mu \lambda \gamma \delta }p_{\gamma }p_{\delta
}^{\prime }p_{\nu }-\epsilon _{_{\nu }\lambda \gamma \delta }p_{\gamma
}p_{\delta }^{\prime }p_{\mu }\right) +\ldots
\end{eqnarray}%
The same correlation function expressed in terms of quark propagators takes
the following form
\begin{eqnarray}
&&\Pi _{\mu \nu \lambda }^{\mathrm{QCD}}(p,q)=\int d^{4}xe^{ipx}\frac{%
\epsilon \widetilde{\epsilon }}{\sqrt{2}}\left[ \gamma _{5}\widetilde{S}%
_{b}^{ib}(x){}\gamma _{5 }\sigma _{\mu \nu }\right.  \notag \\
&&\left. \times \widetilde{S}_{b}^{ei}(-x){}\gamma _{\lambda }+\gamma
_{\lambda }\widetilde{S}_{b}^{ib}(x){}\gamma _{5}\sigma _{\mu \nu }%
\widetilde{S}_{b}^{ei}(-x){}\gamma _{5}\right] _{\alpha \beta }  \notag \\
&&\times \langle \pi (q)|\overline{u}_{\alpha }^{a}(0)d_{\beta
}^{d}(0)|0\rangle .  \label{eq:CorrF8}
\end{eqnarray}%
Expanding $\overline{u}_{\alpha }^{a}d_{\beta }^{d}$ in accordance with Eq.\
(\ref{eq:MatEx}) and substituting into Eq.\ (\ref{eq:CorrF8}) local matrix
elements of the pion we obtain $\Pi _{\mu \nu \lambda }^{\mathrm{QCD}}(p,q)$
which can be matched to $\Pi _{\mu \nu \lambda }^{\mathrm{Phys}}(p,q)$ to
fix same tensor structures. In order to derive sum rule we use structures $%
\sim \epsilon _{\mu \lambda \gamma \delta }p_{\gamma }p_{\delta }^{\prime
}p_{\nu }$ from both sides of equality. The pion matrix element that
contributes to this structure is%
\begin{equation*}
0|\overline{d}(0)\gamma _{5}\gamma _{\mu }u(0)|\pi (q)\rangle =if_{\pi
}q_{\mu }.
\end{equation*}%
In fact, it can be included into the chosen structure after replacement $%
q_{\mu }=p_{\mu }^{\prime }-p_{\mu }$ .

In obtained equality we apply the soft limit $q\rightarrow 0$ ($p=p^{\prime
} $) and perform the Borel transformation on variable $p^{2}$. This
operations leads to a sum rule for two strong couplings $g_{Z_{b}h_{b}\pi }$
and $g_{Z_{b}h_{b}^{\prime }\pi }$. The second expression is obtained from
the first one by applying the operator $d/d(-1/M^{2})$ .

The principal output of these calculations, i.e. the spectral density $\rho
_{\mathrm{h}}^{\mathrm{QCD}}(s)$ reads
\begin{equation}
\rho _{\mathrm{h}}^{\mathrm{QCD}}(s)=\frac{f_{\pi }}{12\sqrt{2}}\left[ \rho
^{\mathrm{pert.}}(s)+\rho ^{\mathrm{n.-pert.}}(s)\right] ,
\end{equation}%
where its perturbative part is given by the formula
\begin{equation}
\rho ^{\mathrm{pert.}}(s)=\frac{(s+2m_{b}^{2})\sqrt{s(s-4m_{b}^{2})}}{\pi
^{2}s^{2}}.
\end{equation}%
The nonperturbative component of $\rho _{\mathrm{h}}^{\mathrm{QCD}}(s)$
includes contributions up to eight dimensions and has the form
\begin{eqnarray}
&&\rho ^{\mathrm{n.-pert.}}(s)=\Big \langle\frac{\alpha _{s}G^{2}}{\pi }\Big
\rangle m_{b}^{2}\int_{0}^{1}f_{1}(z,s)dz  \notag \\
&&+\Big \langle g_{s}^{3}G^{3}\Big \rangle\int_{0}^{1}f_{2}(z,s)dz  \notag \\
&&-\Big \langle\frac{\alpha _{s}G^{2}}{\pi }\Big \rangle^{2}m_{b}^{4}%
\int_{0}^{1}f_{3}(z,s)dz.  \label{eq:NPert}
\end{eqnarray}%
Here the functions $f_{k}(z,s)$ are:%
\begin{equation*}
f_{1}(z,s)=\frac{1}{3}\frac{(1+3r)}{r^{2}}\delta ^{(2)}(s-\Phi ),
\end{equation*}%
\begin{eqnarray*}
&&f_{2}(z,s)=\frac{1}{15\cdot 2^{6}}\frac{1}{r^{4}}\left\{
4r^{2}(3+17r+21r^{2})\delta ^{(2)}(s-\Phi )\right. \\
&&+2r\left[ sr^{2}(4+13r)+3m_{b}^{2}(3+16r+18r^{2})\right] \delta
^{(3)}(s-\Phi ) \\
&&+\left[ s^{2}r^{4}+6m_{b}^{2}sr^{2}(1+3r)-7m_{b}^{4}(1+5r+5r^{2})\right] \\
&&\left. \times \delta ^{(4)}(s-\Phi )\right\} ,
\end{eqnarray*}%
\begin{equation*}
f_{3}(z,s)=\frac{1}{54}\frac{\pi ^{2}}{r^{2}}\delta ^{(5)}(s-\Phi ),
\end{equation*}%
where
\begin{equation*}
\ r=z(z-1),\ \Phi =\frac{m_{b}^{2}}{z(1-z)}.
\end{equation*}%
In the expressions above the Dirac delta function $\ \delta ^{(n)}(s-\Phi )$
is defined in accordance with
\begin{equation}
\delta ^{(n)}(s-\Phi )=\frac{d^{n}}{ds^{n}}\delta (s-\Phi ).
\end{equation}

\begin{widetext}

\begin{figure}[h!]
\begin{center}
\includegraphics[totalheight=6cm,width=8cm]{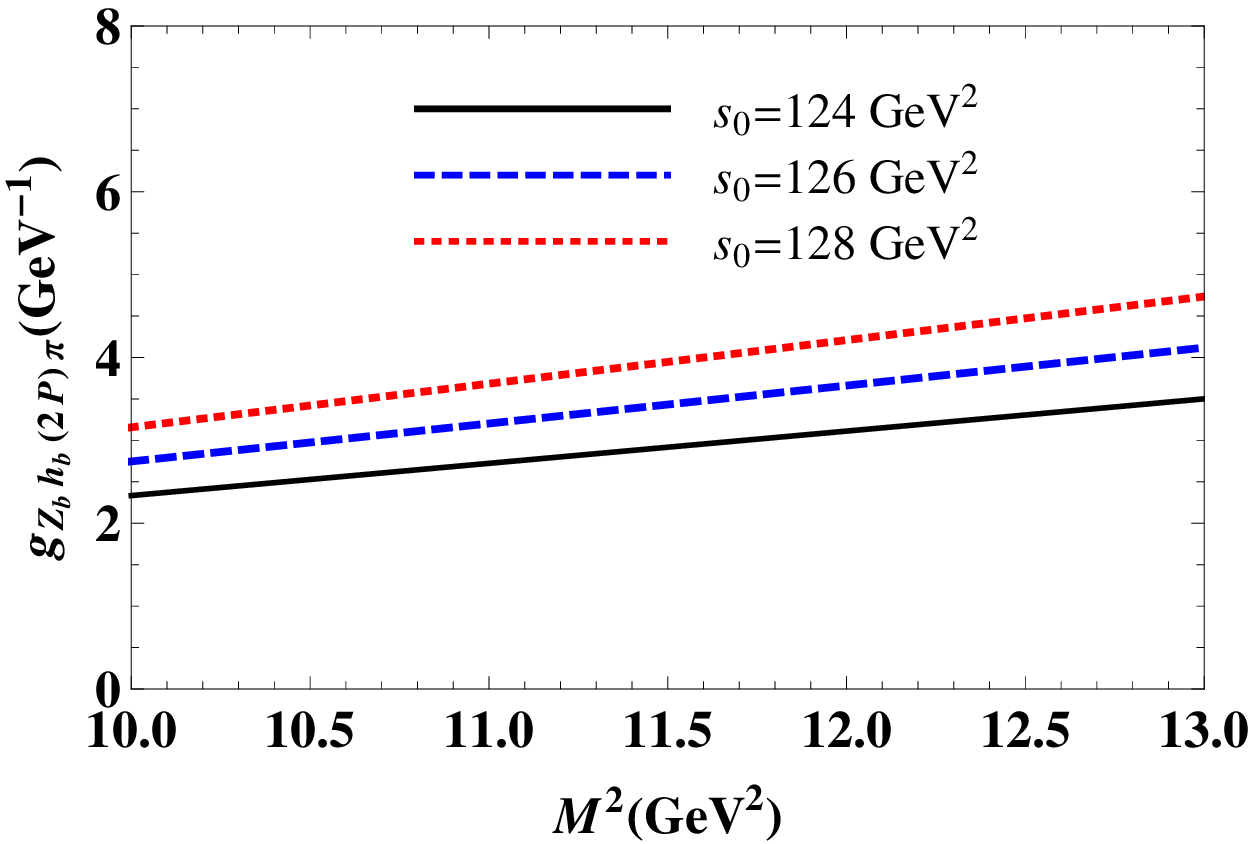}\,\,
\includegraphics[totalheight=6cm,width=8cm]{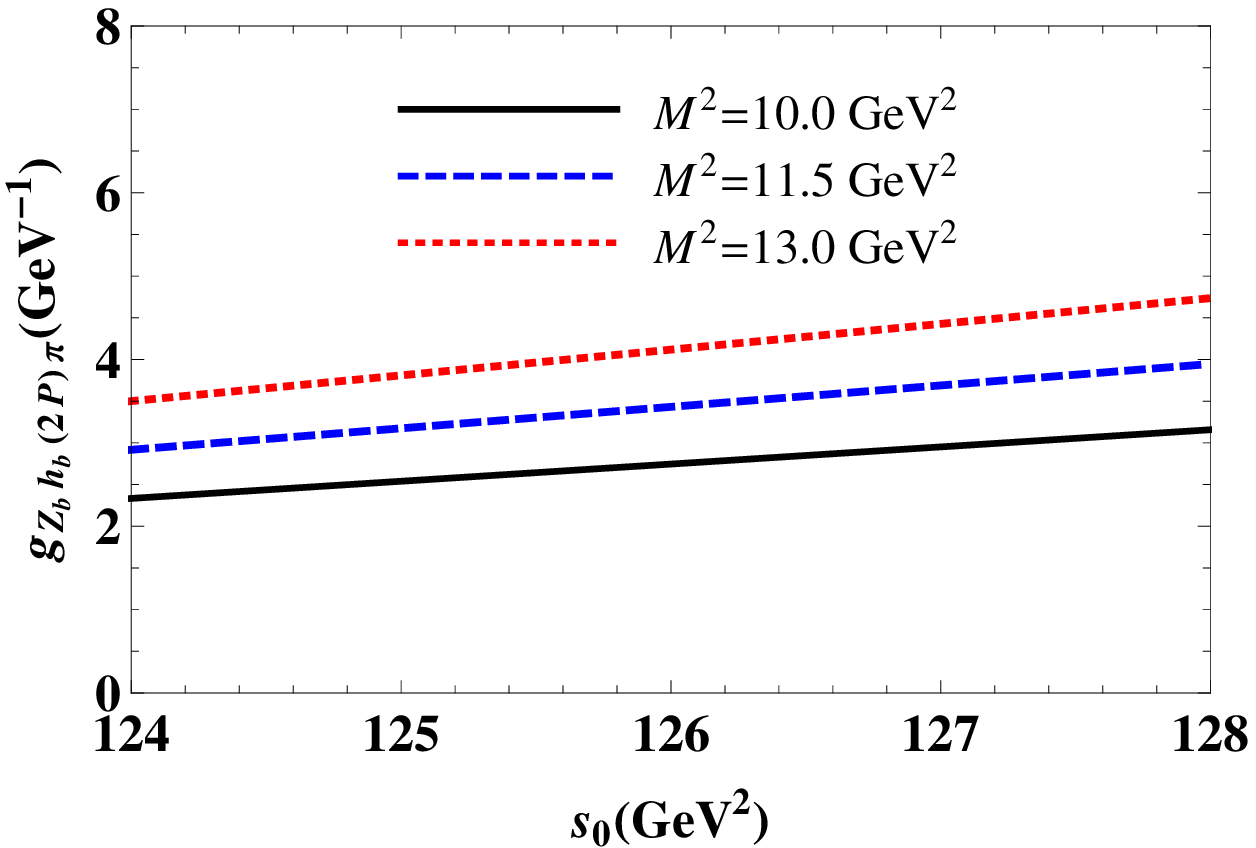}
\end{center}
\caption{ The coupling $g_{Z_bh_{b}^{\prime}\pi}$  vs  Borel parameter $M^2$
(left panel), and  continuum threshold $s_0$ (right panel).}
\label{fig:StrCoup}
\end{figure}

\end{widetext}

The width of the decays $Z_{b}\rightarrow h_{b}(1P)\pi $ and $%
Z_{b}\rightarrow h_{b}(2P)\pi $ are calculated using the formula%
\begin{equation*}
\Gamma (Z_{b}\rightarrow h_{b}(mP)\pi )=g_{Z_{b}h_{b}(mP)\pi }^{2}\frac{%
\lambda \left( m_{Z_{b}},\ m_{h(mP)},m_{\pi }\right) ^{3}}{12\pi }.
\end{equation*}%
In numerical computations we employ parameters of the $h_{b}(mP)$ mesons
obtained in the previous subsection. The working regions of the Borel
parameter $M^{2}$ and continuum threshold $s_{0}$ are the same as in
analysis of $Z_{b}\rightarrow \Upsilon (nS)\pi $ decays. Below we provide
our results for the strong couplings (in units $\mathrm{GeV}^{-1}$)
\begin{equation}
g_{Z_{b}h_{b}\pi }=0.94\pm 0.27,\ \,g_{Z_{b}h_{b}^{\prime }\pi }=3.43\pm
0.93.  \label{eq:Couplh}
\end{equation}%
In Fig.\ \ref{fig:StrCoup} we plot the coupling $g_{Z_bh_{b}^{\prime}\pi}$
as a function of the Borel parameter and continuum threshold to show its
dependence on these auxiliary parameters. It is easy to see that theoretical
errors are within limits accepted in sum rule calculations.

Using Eq.\ (\ref{eq:Couplh}) it is not difficult we evaluate width of the
decays:
\begin{eqnarray}
\Gamma (Z_{b} &\rightarrow &h_{b}(1P)\pi )=6.30\pm 1.76\ \mathrm{MeV,}
\notag \\
\Gamma (Z_{b} &\rightarrow &h_{b}(2P)\pi )=7.35\pm 2.13\ \mathrm{MeV.}
\label{eq:ResultsDW2}
\end{eqnarray}

\section{Analysis and concluding notes}

\label{sec:Analysis}


The experimental data on decay channels of the $Z_{b}(10610)$ resonance were
studied and presented in a rather detailed form in Refs.\ \cite%
{Belle:2011aa,Garmash:2014dhx,Garmash:2015rfd}. Its full width was estimated
as $\Gamma =18.4\pm 2.4\ \mathrm{MeV}$ essential part of which, i.e.
approximately $86\%$ of $\Gamma $ is due to decay $Z_{b}\rightarrow B^{+}%
\overline{B}^{\ast 0}+B^{\ast +}\overline{B}^{0}$. The remaining part of the
full width is formed by five decay channels investigated in the present
work. It is clear that our results for width of decays $Z_{b}\rightarrow
\Upsilon (nS)\pi $ and $Z_{b}\rightarrow h_{b}(nP)\pi $ overshoot the
experimental data. Therefore, in the light of present studies we refain from
interpretation of the $Z_{b}(10610)$ resonance as a pure diquark-antidiquark
$[bu][\overline{b}\overline{d}]$ state.

Nevertheless, encouraging are theoretical predictions for the ratios
\begin{equation}
\mathcal{R(}n)\mathcal{=}\frac{\Gamma (Z_{b}\rightarrow \Upsilon (nS)\pi )}{%
\Gamma (Z_{b}\rightarrow \Upsilon (1S)\pi )},\mathcal{R(}m)\mathcal{=}\frac{%
\Gamma (Z_{b}\rightarrow h_{b}(mS)\pi )}{\Gamma (Z_{b}\rightarrow \Upsilon
(1S)\pi )},
\end{equation}%
where we normalize widths of different decay channels to $\Gamma
(Z_{b}\rightarrow \Upsilon (1S)\pi )$. The ratio $\mathcal{R}$ can be
extracted from available experimental data and calculated from decay widths
obtained in the present work. In order to fix existing similarities and
differences between theoretical and experimental information on $\mathcal{R}$
we provide two sets of corresponding values in Table\ \ref{tab:BRatio}. It
is worth to note that we use latest available experimental information from
Ref. \cite{Olive:2016xmw}.

It is seen that theoretical predictions follow pattern of experimental data:
we observe the same hierarchy of theoretical and experimental decay widths.
At the time, numerical differences between them are noticeable.
Nevertheless, in a result of large errors in both sets, there are sizeable
overlap regions for each pair of $\mathcal{R}$s, which demonstrate not only
qualitative agreement between them but also quantitative compatibility of
two sets.

These observations may help one to understand the nature of the $Z_{b}$
resonance. The Belle Collaboration discovered two $Z_{b}$ and $Z_{b}^{\prime
}$ resonances with very close masses. We have calculated parameters of an
axial-vector diquark-antidiquark state $[bu][\overline{b}\overline{d}]$, and
interpreted it as $Z_{b}$. It is possible to model the second $Z_{b}^{\prime
}$ resonance using alternative interpolating current, as it has been
emphasized in Sec. \ref{sec:MassCoupl} and explore its properties. The
current with the same quantum numbers but different color organization may
also play a role of such alternative (see, for example, Ref.\ \cite%
{Agaev:2017foq}). One of possible scenarios implies that observed resonances
are admixtures of these tetraquarks, which may fit measured decay widths.

The diquark-antidiquark interpolating current used in the present work can
be rewritten as a sum of molecular-type terms. In other words, some of
molecular-type currents effectively contribute to our predictions, and by
enhancing these components (i.e. by adding them to interpolating current
with some coefficients) better agreement with experimental data may be
achieved. In other words, the resonances $Z_{b}$ and $Z_{b}^{\prime }$ may
"contain" both the diquark-antidiquark and molecular components.

Finally, $Z_{b}$ and $Z_{b}^{\prime }$ states may have pure molecular
structures. But pure molecular-type bound states of mesons are usually
broader than diquark-antidiquarks with the same quantum numbers and quark
contents. In any case, all these suggestions require additional and detailed
investigations.

In the present study we have fulfilled only a part of this program. In the
framework of QCD sum rule methods we have calculated the spectroscopic
parameters of $Z_{b}$ state by modeling it as diquark-antidiquark state, and
found widths five of its observed decay channels. We have also evaluated
mass and decay constant of $h_{b}(2P)$ meson, which are necessary for
analysis of $Z_{b}\rightarrow h_{b}(2P)\pi $ decay. Calculation of the $%
Z_{b} $ resonance's dominant decay channel may be performed, for example,
using QCD three-point sum rule approach, which is beyond the scope of the
present work. Decays considered here involve excited mesons $\Upsilon (nS)$
and $h(mP)$, parameters of which require detailed analysis in a future. More
precise measurements of $Z_{b}$ and $Z_{b}^{\prime }$ partial decays' width
can also help in making a choice between outlined scenarios.
\begin{table}[tbp]
\begin{tabular}{|c|c|c|c|c|}
\hline\hline
$\mathcal{R}$ & $n=2$ & $n=3$ & $m=1$ & $m=2$ \\ \hline\hline
Exp. \cite{Olive:2016xmw} & $6.67_{-2.37}^{+3.11}$ & $3.89_{-1.55}^{+2.02}$
& $6.48_{-2.45}^{+3.18}$ & $8.70_{-3.41}^{+4.39}$ \\ \hline
This work & $12.63 \pm 5.43$ & $6.08 \pm 2.76$ & $4.63 \pm 1.95$ & $5.40 \pm
2.32$ \\ \hline\hline
\end{tabular}%
\caption{Experimental values and theoretical predictions for $\mathcal{R}$.}
\label{tab:BRatio}
\end{table}


\section*{ACKNOWLEDGEMENTS}

S.~S.~A. thanks T.~M.~Aliev for helpful discussions. K.~A.~ thanks T\"{U}%
BITAK for the partial financial support provided under Grant No. 115F183.

\end{document}